\documentclass[preprint,11pt,oneside]{article}
\usepackage[english]{babel}
\usepackage[T1]{fontenc}
\usepackage{graphicx}
\usepackage{bm}
\usepackage{amsmath,amsfonts,amssymb,amsthm}
\usepackage{latexsym}
\usepackage{booktabs}
\usepackage{multirow}
\usepackage[TABBOTCAP]{subfigure}
\usepackage{cases}
\usepackage{bm}

\newcommand{\beq}{\begin{equation}}
\newcommand{\eeq}{\end{equation}}

\def\be{\begin{equation}}
\def\ee{\end{equation}}

\theoremstyle{remark}

\usepackage{xcolor}
\definecolor{blun}{cmyk}{0.8, 0.5, 0, 0.7}
\definecolor{darkpastelred}{rgb}{0.76, 0.23, 0.13}
\definecolor{bostonuniversityred}{rgb}{0.8, 0.0, 0.0}

\usepackage[bookmarks,colorlinks,pdfauthor={Cinzia Soresina},linkcolor=blun,citecolor=blun]{hyperref}

\providecommand{\keywords}[1]{\textbf{\textit{Keywords:}} #1}

\usepackage[author={Cinzia}]{pdfcomment}

\definecolor{inchworm}{rgb}{0.7, 0.93, 0.36}

\begin{document}

\let\oldproofname=\proofname
\renewcommand{\proofname}{\rm\bf{\oldproofname}}

\title{\Large{The effects of fecundity, mortality and distribution of the initial condition in phenological models}}
\author{\normalsize{S. Pasquali}\\ \footnotesize{CNR-IMATI ``Enrico Magenes'', via Alfonso Corti 12, 20133 Milano, Italy}\\[0.2cm]
\normalsize{C. Soresina\footnote{corresponding author: soresina@ma.tum.de}}\\ \footnotesize{Department of Mathematics - Technical University of Munich}\\[-0.1cm] \footnotesize{Boltzmannstr. 3, 85748 Garching bei M\"unchen, Germany}\\[0.2cm]
\normalsize{G. Gilioli} \\ \footnotesize{Department of Molecular and Translational Medicine, University of Brescia,}\\[-0.1cm] \footnotesize{Viale Europa 11, 25123 Brescia, Italy}}


\maketitle

\begin{abstract}
Pest phenological models describe the cumulative flux of the individuals into each stage of the life cycle of a stage-structured population. Phenological models are widely used tools in pest control decision making. Despite the fact that these models do not provide information on population abundance, they share some advantages with respect to the more sophisticated and complex demographic models. The main advantage is that they do not require data collection to define the initial conditions of model simulation, reducing the effort for field sampling and the high uncertainty affecting sample estimates. Phenological models are often built considering the developmental rate function only. To the aim of adding more realism to phenological models, in this paper we explore the consequences of improving these models taking into consideration three additional elements: the age distribution of individuals which exit from the overwintering phase, the age- and temperature-dependent profile of the fecundity rate function and the consideration of a temperature-dependent mortality rate function. Numerical simulations are performed to investigate the effect of these elements with respect to phenological models considering development rate functions only. To further test the implications of different models formulation, we compare results obtained from different phenological models to the case study of the codling moth (\emph{Cydia pomonella}) a primary pest of the apple orchard. The results obtained from model comparison are discussed in view of their potential application in pest control decision support.
\end{abstract}



\section{Introduction}

Understanding the phenology of pests under field conditions and the ecological factors influencing development, survival, and reproduction rates determining the change in pest population abundance are key issues in developing strategies for pest management. To this end, mathematical models of stage-structured population are suitable tools for the description of complex processes at the individual level (i.e., the life cycle strategies) and to predict their consequences in terms of population dynamics \cite{ dicola1998,dicola1999,gutierrez1996,metz1986}. Reliable models can support decision making in pest management tactics and strategies to improve the effectiveness of pest control decisions. The capability to predict population dynamics and evaluate scenarios of pest control in agro-ecosystems under a variety of environmental conditions can reduce the number and cost of control intervention, improving crop yield and quality as well as the health and sustainability of crop production. 

Plant pests, including insects, mites and nematodes, cannot internally regulate their body temperature, therefore phenological events, as well as the change in physiological age, are dependent on ambient temperatures to which they are exposed. These poikilotherm organisms require a certain amount of heat to 
develop. Measuring the amount of heat accumulated over time provides a physiological time scale that is biologically more accurate than chronological time \cite{dicola1999}. For poikilotherm organisms temperature is also considered the main driving variable for mortality and fecundity. Dependency from other environmental driving (e.g., the influence of relative humidity on survival) and control (e.g., the effect of strong rain or wind in the oviposition behavior) variables can also be considered in poikilotherm population models \cite{gilioli2016,Schmidt2003}.

Stage-structured demographic models have been considered for their capability to describe the temporal dynamics of population abundance and support decision making in pest management. A wide review of recent publications on stage structured population models can be found in \cite{robertson2018}. These demographic models have been applied in literature to describe the population dynamics of plant pests, some examples can be found in \cite{blum2018,gilioli2016,gilioli2014, langille2016,lu2017}. 
More recently they have also been used in pest risk assessment of invasive alien species and to comparative evaluate risk scenarios and the efficacy of risk reducing options \cite{edholm2018,pasquali2015,sporleder2009}.

Phenological models are by far the most widely used tools in pest control decision support. They are often stage-structured models and predict the time of significant events in an organism development through the cumulative flux of the individuals into each stage (in terms of percentage of development completion for each stage). However they do not consider population abundance.

Both demographic and phenological temperature-dependent structured population models can be described by systems of partial differential equations (PDEs) \cite{BuffoniPasquali2007, dicola1998, dicola1999, gutierrez1996, metz1986}. In these models the influence of temperature on the components of the life history strategies (e.g., development, mortality and fecundity) is described by temperature-dependent rate functions.

Stage structured demographic models are powerful tools able to describe the change in population abundance both in time and age. This provides opportunity for interpreting the impact of pest population on both natural and cultivated plants, since population abundance is the major driving force acting on the host plants \cite{gilioli2014}. However, to obtain information on the abundance in stage-structured models, the definition of the initial conditions in terms of distribution of individuals between and within the stages is needed. Obtaining this information often requires a monitoring effort that exceeds the resources available in many Integrated Pest Management (IPM) programs. Furthermore, the reliability of population density estimation is also a factor highly impacting on the uncertainty associated to model output.

Phenological model instead, share some advantages with respect to the more sophisticated and complex demographic models. They can be computed starting from a fixed initial condition (usually the 100\% of individuals in the overwintering stage at a specific point in time, e.g. January $1^{st}$), and the phenological dynamics can be derived, in the simplest case, considering stage-specific development rate functions only and a time series of temperature. The recruitment is also present in the model, but it is expressed in terms of the adult development and allows the production of a single egg for each adult, so keeping constant the number of individuals in time. From now on, this simple model based only on development rate functions, and with initial condition of overwintering individuals having physiological age zero, will be denoted by M0. 

However, the advantages offered by simple phenological models in terms of easy of parameterization should be considered together potential limitations deriving from the use of development rate functions only. In this paper we explore these limitations comparing the performance of M0 with a more complete formulation of the phenological model which includes element of biological realism. To this aim we introduce three different alternative formulations of the model M0 (that can also be combined together):
\begin{itemize}
\item[-] Phenological model M1. This formulation accounts for the age distribution of individuals which exit the overwintering phase. We assume this distribution is kept also in the individuals emerging from the diapause period when temperature and other environmental conditions trigger the development process. To the best of our knowledge, the age distribution at the beginning of the development is usually disregarded in phenological models but it can potentially have important influence on population phenology. 
\item[-] Phenological model M2. In this formulation the fecundity is introduced by considering various oviposition rates and profiles modifying the input flux in the eggs stage. The fecundity is function of both the temperature and the age of the adult female. The influence of temperature is described by a parabolic function, widely used in literature on modelling oviposition rate \cite{mishra2004, sporleder2004, damos2008, gutierrez2012, gilioli2016}. To account for the  adult age, we compare various fecundity profiles ranging from adults immediately reproductive after emergence with a peak of oviposition in the first part of their life to reproductive profiles characterized by  a pre-oviposition period and a peak of oviposition late in the adult stage.
\item[-] Phenological model M3. In this formulation the mortality is introduced considering a temperature-dependent mortality rate function characterized by a minimum in the range of optimal temperature and a bath-tube profile \cite{wang2002}. This pattern is common in poikilotherm organisms with low mortality values in a suitable temperature interval and increasing mortality outside the interval, for higher and lower temperatures.
\end{itemize}
The four model formulations were analyzed numerically to explore the effects of fecundity, mortality, and distribution of the initial condition over the physiological age, on the pest phenology. Then, we consider an application to a specific pest, the codling moth, for which data on adult dynamics have been collected in a specific location of Northern Italy. The comparison with field data allows to best point out the differences of the various formulation of the phenological model.

The paper is organized as follow: in Section \ref{MatMod} a continuous-time model of stage-structured population dynamics is presented; models M0, M1, M2, and M3 are compared in Section \ref{ConfrMod} using a set of general biodemographic functions, realistic for the biology of pests, but not calibrated for a specific pest; in Section \ref{cydia}, an application to the codling moth is considered. Finally, in Section \ref{Concl} some concluding remarks can be found.

\section{The mathematical model}\label{MatMod} 

The phenological model is based on a system of partial differential equations that allows to obtain the temporal dynamics of a stage-structured population and the distribution of the individuals on physiological age within each stage. Let 
$$\phi^i(t,x) dx = \textnormal{ number of individuals in stage $i$ at time $t$ with age in $(x,x+dx)$,}$$
$i=1,2,...,s$, where $s$ is the number of stages. Stages from $1$ to $s-1$ are immature stages, and stage $s$ represents the reproductive stage (adult individuals). The variable $t$ denotes the chronological time while $x$ is a developmental index which represents the physiological age indicating the development over time \cite{BuffoniPasquali2007, BuffoniPasquali2010,BuffoniPasquali2013,dicola1999}. The functions $\phi^i(t,x)$ allow to obtain the number of individuals in stage $i$ at time $t$:
$$N^i(t)=\int_0^1 \phi^i(t,x) dx.$$

We consider a stochastic approach which allows to take into account the variability of the development rate among the individuals \cite{BuffoniPasquali2010,BuffoniPasquali2013}. The dynamics is described in terms of the forward Kolmogorov equations \cite{gardiner1986,carpi1988}

\beq
\frac {\partial \phi^i} {\partial t} + \frac {\partial} {\partial x}
\left[ v^i(t) \phi^i - \sigma^i \frac  {\partial \phi^i} {\partial x} \right]+ m^i(t) \phi^i
 = 0, \quad t > t_0, \; x \in (0,1),
\label{kolm1}
\eeq

\beq
\left[ v^i(t) \phi^i(t,x) - \sigma^i \frac {\partial \phi^i} {\partial x}
\right]_{x=0} = F^i(t),
\label{kolm2}
\eeq

\beq
\left[ - \sigma^i \frac {\partial \phi^i} {\partial x} \right]_{x=1} = 0,
\label{kolm3}
\eeq

\beq
\phi^i(t_0,x) = {\hat \phi}^i(x),
\label{kolm4}
\eeq
where $i=1,2,...s$, $v^i(t)$ and $m^i(t)$ are the specific development and mortality rates assumed independent of the age $x$, ${\hat \phi}^i(x)$ are the initial distributions, while $\sigma^i$ are the diffusion coefficients, assumed time independent. The boundary condition at $x=0$ assigns the input flux into stage $i$, while the boundary condition at $x=1$ means that the output flux from stage $i$ is due only to the advective component $v^i(t) \phi^i(t,1)$ \cite{BuffoniPasquali2007}. The terms $F^i(t)$, when $i>1$, are the individual fluxes from stage $i-1$ to stage $i$ and are  
\beq
F^i(t) = v^{i-1}(t) \phi^{i-1}(t,1), \quad i>1,
\label{vonfos5}
\eeq
while the term $F^1(t)$ is the eggs production flux. We consider three different formulation of the reproduction term:
\begin{itemize}
\item
Fecundity dependent on the adult temperature-dependent development. It can be either model M0 or M1, where the term $F^1(t)$ is given by
\beq
F^1(t) = v^s(t). 
\label{Fclassico}
\eeq
The adults produce eggs through their development and each adult gives rise to a single egg during the whole physiological life. This results in a constant input flux in each stage. It is worthwhile to note that this formulation does not take into account a different eggs production with respect to the physiological age.

\item 
Fecundity dependent on physiological age. In this case (which can be included in model M2) we consider the age-dependent flux 
\beq
F^1(t) = \int_0^1 \; f(x) \; \phi^s(t,x) \; dx,
\label{Fmod_x}
\eeq
where $f$ describes the oviposition profile with respect to the physiological age. Also with this formulation it is possible to rescale to one the number of eggs produced by each adult female. Note that in this case no temperature dependence is present. However, it is known that the eggs production is influenced by temperature, and in a different way with respect to (\ref{Fclassico}). For this reason, we want to deal with a flux which depends on both temperature and physiological age.
\item 
Fecundity dependent on physiological age and on temperature. In this case (also included in model M2) $F^1(t)$ is given by
\beq
F^1(t) = b(T(t))\int_0^1 \; f(x) \; \phi^s(t,x) \; dx,
\label{Fmod_xT}
\eeq
where $b(T(t))$ is the temperature function depending on the chronological time trough the temperature $T$. With this choice, we are able to describe a fecundity rate which varies with both temperature and physiological age. 
\end{itemize} 
In the two last cases (model M2), choosing appropriately the profile of the function $f$ it is possible to account for a pre-oviposition period in the fecundity function avoiding the introduction of a further stage that increases the number of differential equations and then the complexity of system \eqref{kolm1}.


\section{Analysis of models M0, M1, M2, M3}\label{ConfrMod}
In this section we explore the effects on the outcomes of different formulation of the phenological model. 
For the purpose of exploring the effects of fecundity, mortality, and age distribution of the initial condition in phenological models, we refer to a generic poikilotherm species characterized by a stage-structured population dynamics dependent on three biodemographic functions (development, mortality and fecundity) and composed by $s=4$ stages: eggs ($i=1$), larvae ($i=2$), pupae ($i=3$) and adults ($i=4$). To run the model we need to specify some parameters and functions involved in system \eqref{kolm1}--\eqref{kolm4}. In next subsection, we explicit a functional form for each biodemographic  function. Then, we will present the comparison of the four models using these biodemographic functions.

\subsection{Biodemographic functions and age distribution of the initial condition in physiological models} \label{CaseStudy}
Biodemographic functions reported in this section are not calibrated on data of a particular species, but they are comparable with those of a insect population. 

\subsubsection{Development}
For all the stages, we consider a Bri\`ere-1 function \cite{Briere1999}, already used to describe the development rate function of pest populations \cite{legaspi2011,falzoi2014,gilioli2017}
\begin{equation}
v(t)=\begin{cases}a T (T-T_m)\sqrt{T_M-T}, & T_m\leq T \leq T_M\\ 0, & \textnormal{otherwise.}\end{cases}
\label{eq:Briere1}
\end{equation}
In \eqref{eq:Briere1}, $a$ is an empirical constant, $T_m$ and $T_M$ are the lower and the lethal temperature thresholds. This nonlinear model involves three parameters and it reproduces a sharp decline above the optimal temperatures, an asymmetry about the optimal temperatures and an inflection point.
Other analytical forms for the development rate function can be found in \cite{Kontodimas2004,quinn2017}.

\begin{figure}%
\centering
  \subfigure[Eggs]{\includegraphics[width=0.35\textwidth]{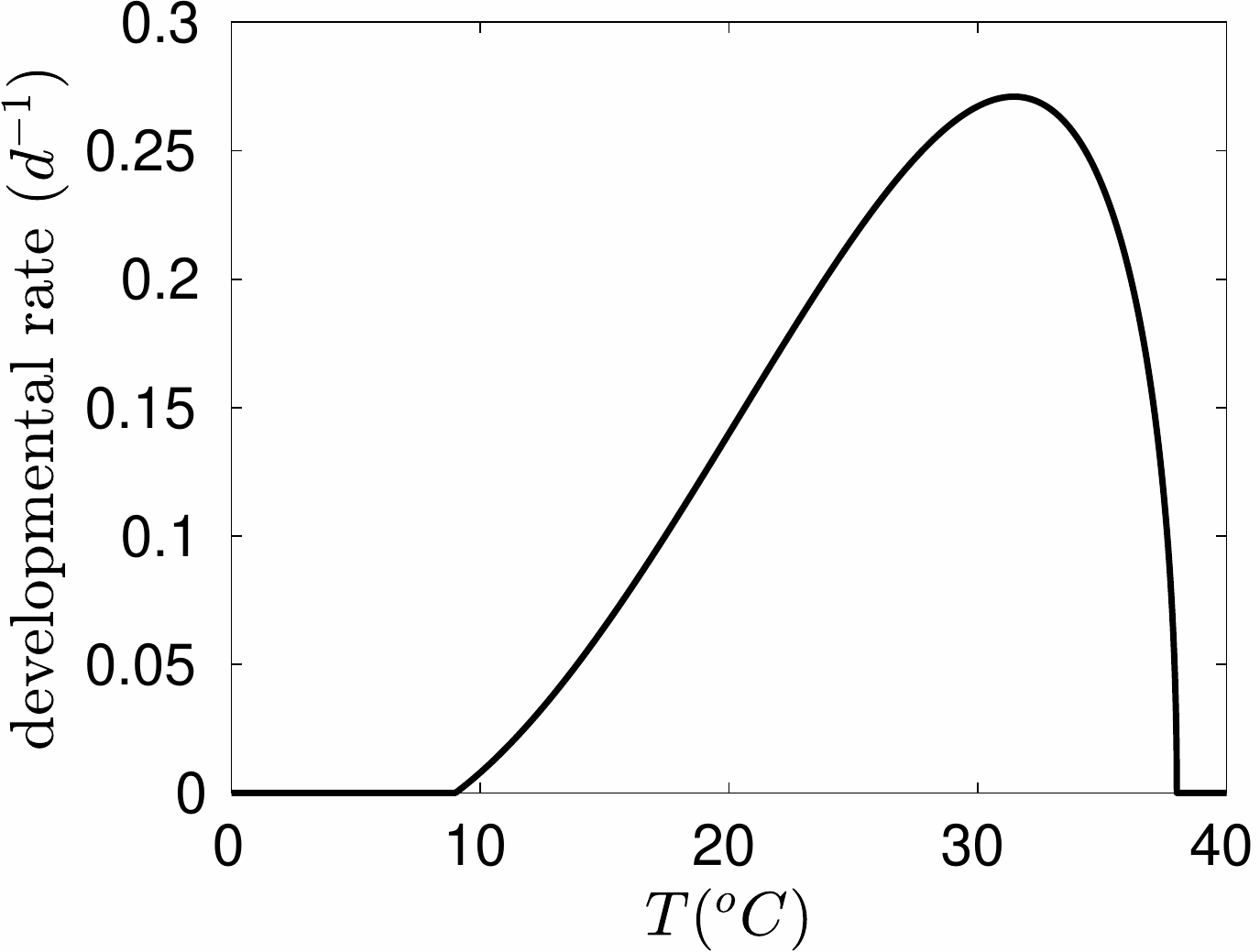}}
	\subfigure[Larvae]{\includegraphics[width=0.35\textwidth]{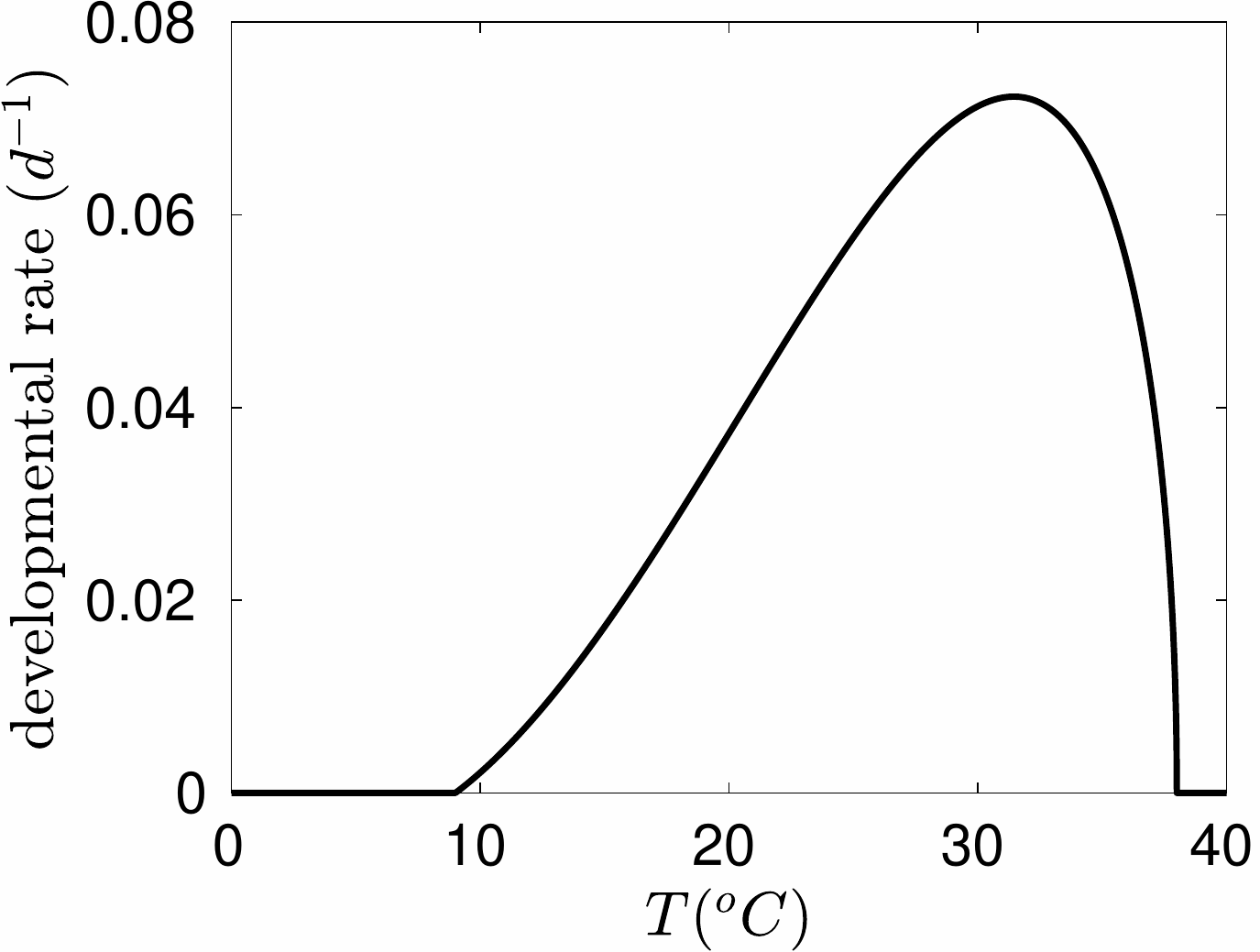}}
	\subfigure[Pupae]{\includegraphics[width=0.35\textwidth]{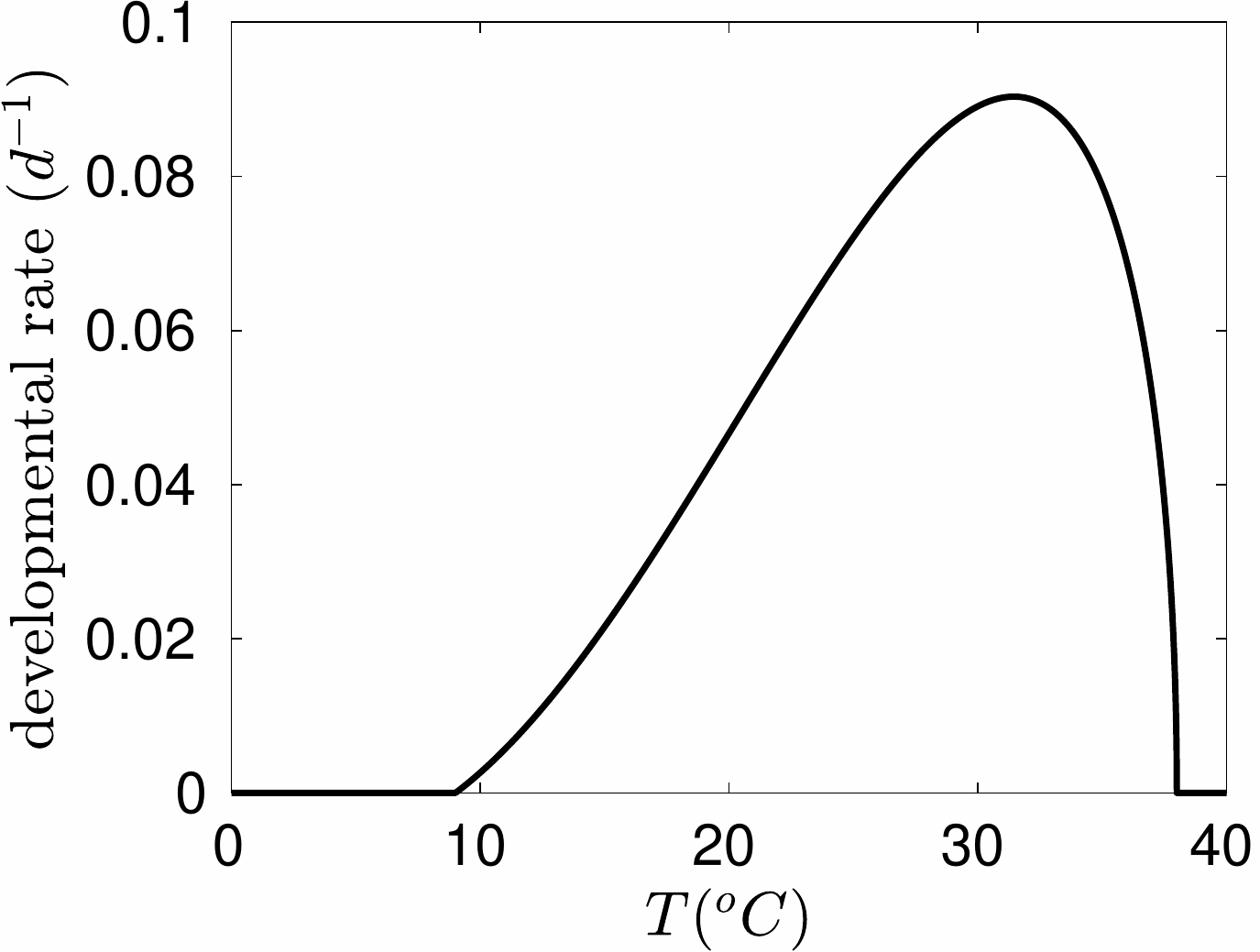}}
	\subfigure[Adults]{\includegraphics[width=0.35\textwidth]{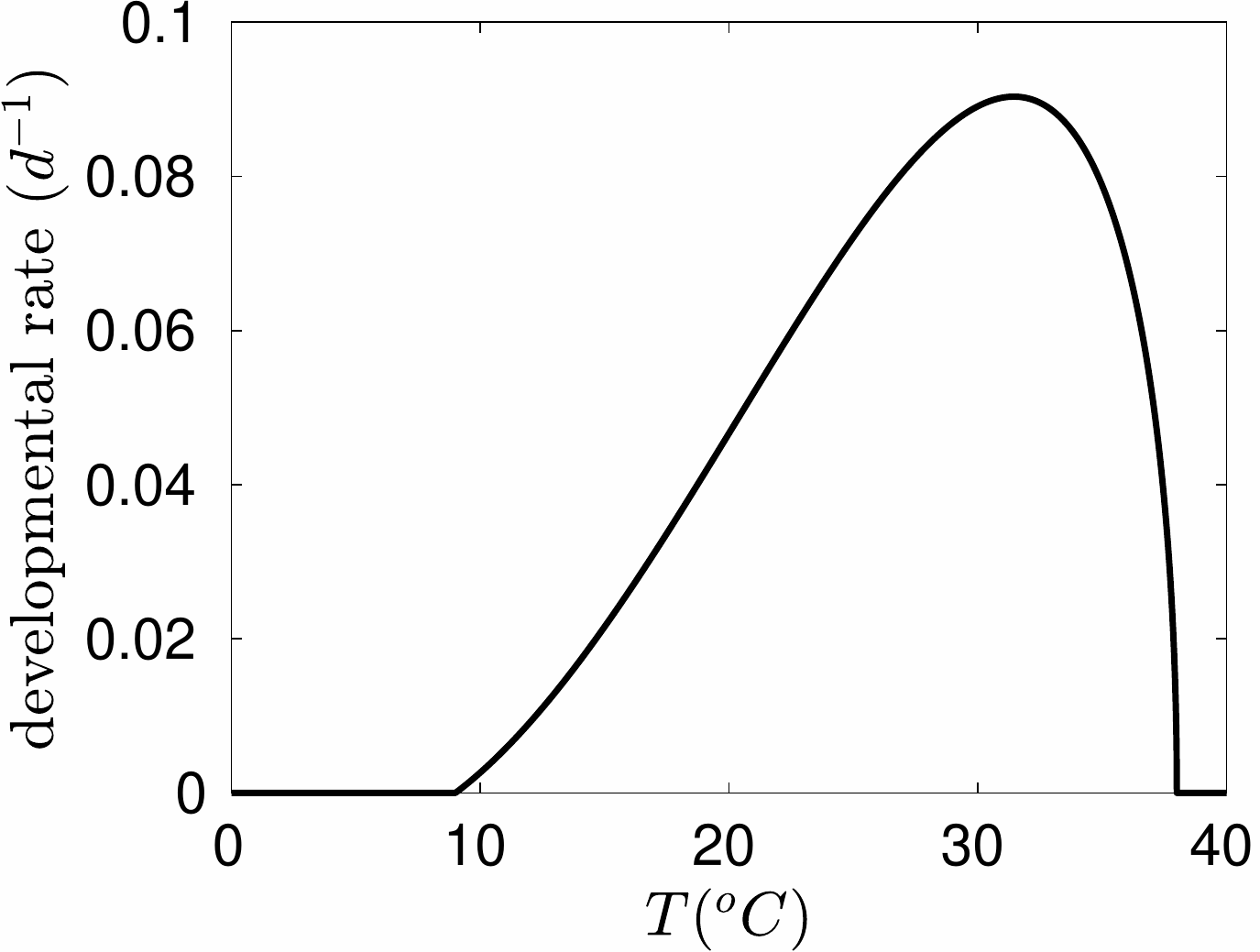}}
\caption{Development rate ($1/days$) as function of temperature ($^\circ C$) for the four biological stages of the population. }
\label{FigDev}
\end{figure}

In the numerical simulations, the values of the parameters reported in Table \ref{TabDev}, for equation \eqref{eq:Briere1}, has been considered. The corresponding development rate functions for the four stages as function of temperature are illustrated in Figure \ref{FigDev}.

\begin{table}[!h]
\centering
\begin{tabular}{ccccc}
\toprule
	    & $i=1$ & $i=2$ & $i=3$ & $i=4$ \\
\midrule
$a$ & $1.5\cdot 10^{-4}$ & $4\cdot 10^{-5}$ & $5\cdot 10^{-5}$ & $5 \cdot 10^{-5}$\\
$T_m\; ( ^\circ C)$ & $9$               & $9$               & $9$              & $9$               \\
$T_M\; ( ^\circ C)$ & $38$              & $38$              & $38$             & $38$              \\
\bottomrule
\end{tabular}
\caption{Parameters of the stage-specific development rate function in \eqref{eq:Briere1} for the four stages: 
 eggs ($i=1$), larvae ($i=2$), pupae ($i=3$) and adults ($i=4$).}
\label{TabDev}
\end{table}

\subsubsection{Fecundity}\label{fec}
Fecundity can be function of either temperature or physiological age or both. Here we consider different oviposition profiles with respect to the physiological age of an adult. To express analytically the fecundity as function of the physiological age, we need to know some quantities strictly connected to the shape of the fecundity rate function: the age $x_*$ at which the pest has the maximum  eggs production, and the physiological age $x_e$ at which an adult has already produced the $99\%$ of the mean total number of eggs $n_e$  laid by an individual at a fixed temperature.

\begin{figure}%
  \subfigure[\label{op02}$x_*=0.2,\;x_e=0.6$]{\includegraphics[width=0.32\textwidth]{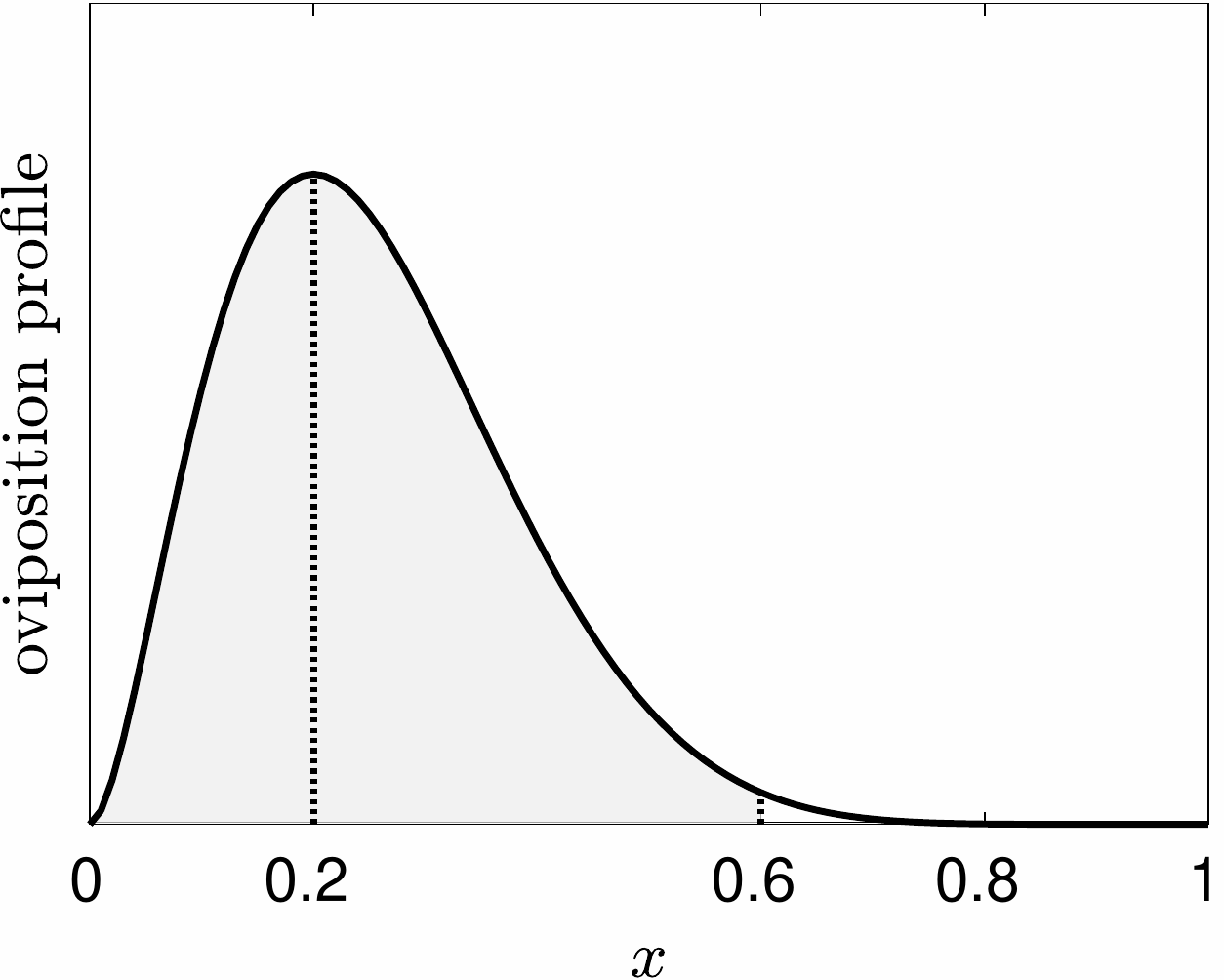}}
	\subfigure[\label{op06}$x_*=0.6,\;x_e=0.8$]{\includegraphics[width=0.32\textwidth]{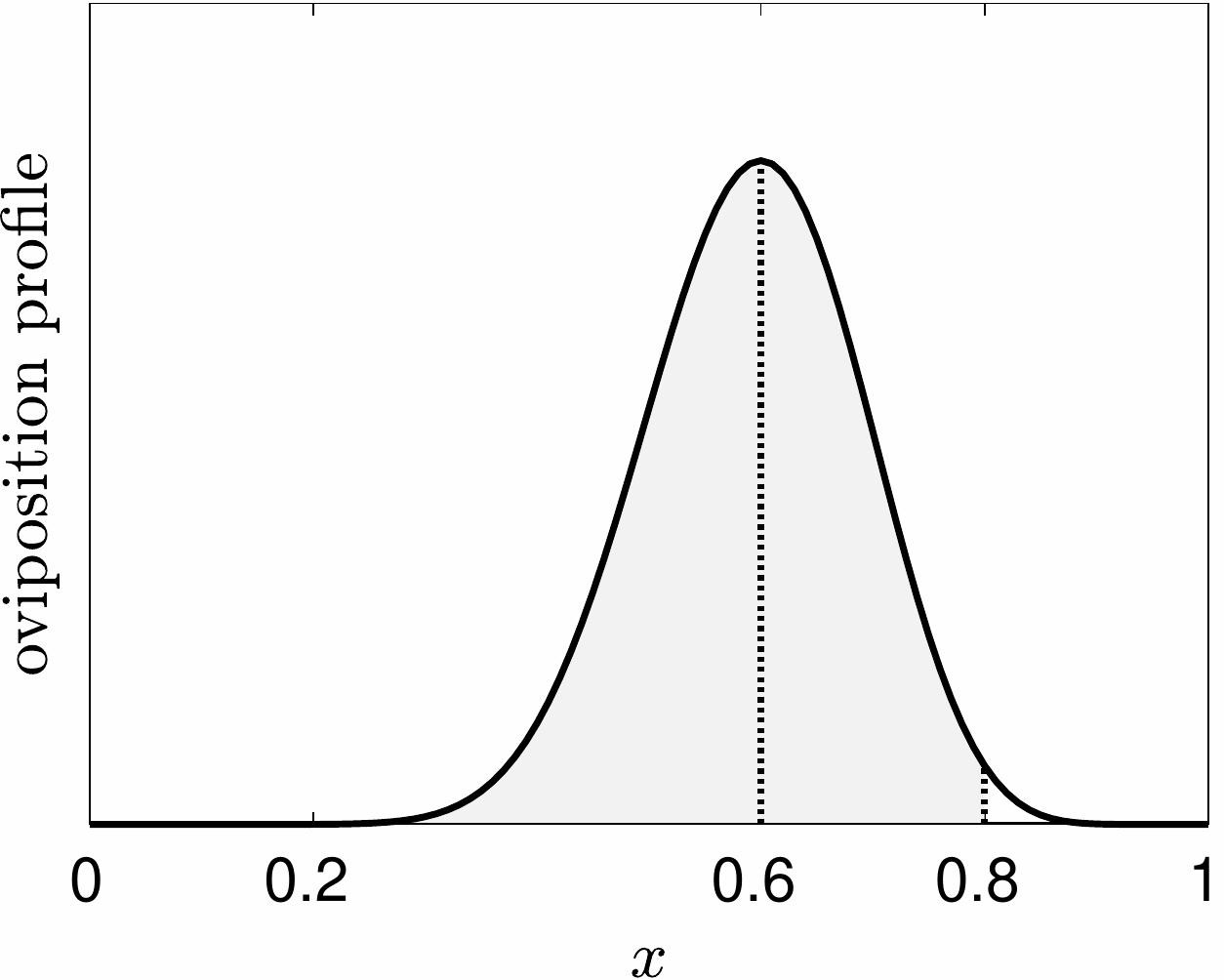}}
	\subfigure[\label{op08}$x_*=0.8,\;x_e=0.95$]{\includegraphics[width=0.32\textwidth]{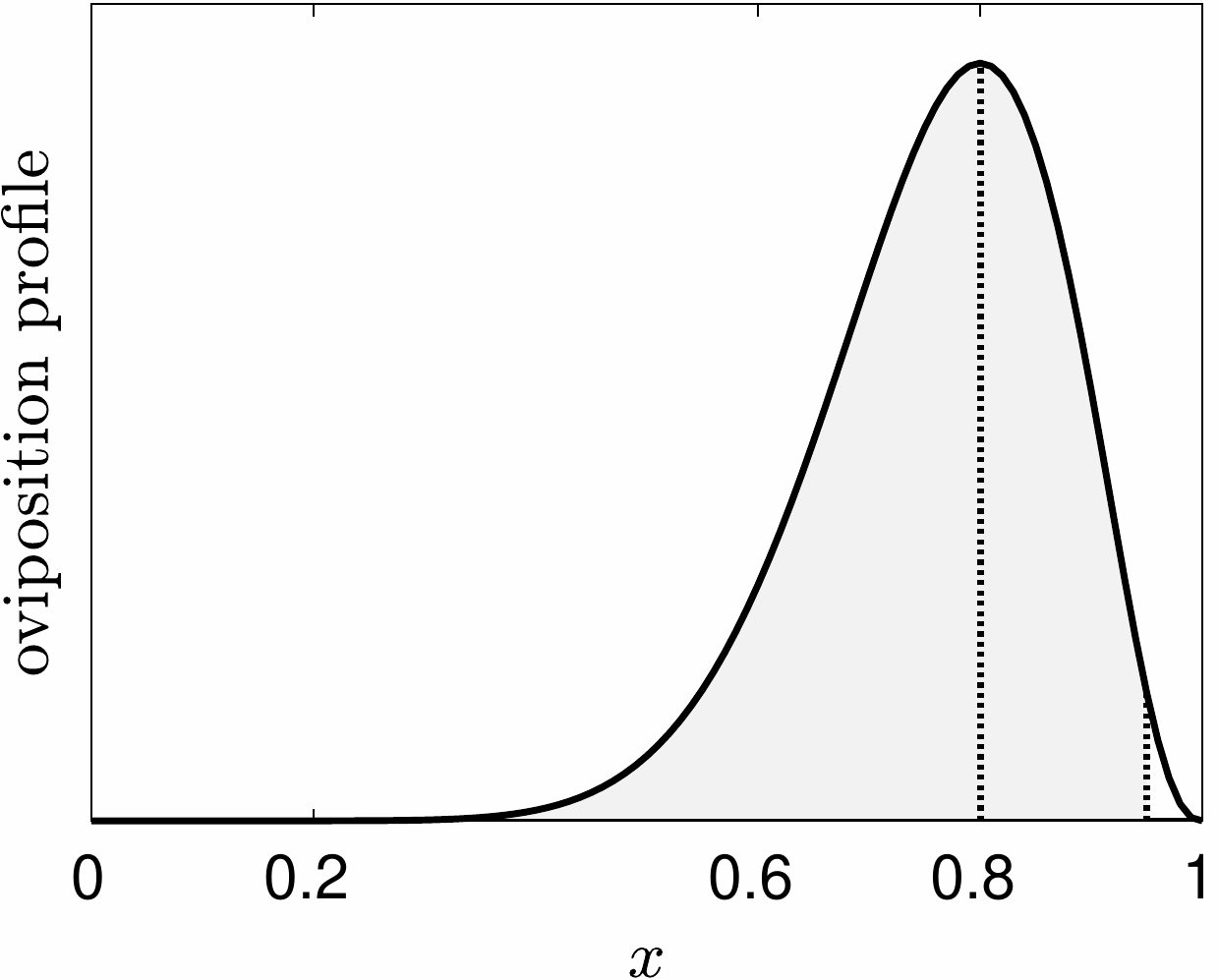}}
\caption{Different oviposition profiles (qualitative).}
\label{FigOvProf}
\end{figure}

We consider here the function $f$ appearing in \eqref{Fmod_x} and \eqref{Fmod_xT}, proportional to a beta density function
\begin{equation}
f(x)=\alpha x^\beta(1-x)^\gamma,
\label{fx}
\end{equation}
where the parameters $\alpha,\;\beta,\,\gamma$ are estimated such that
\begin{equation}\label{vincoli_f}
f(x_*)\geq f(x) \;\forall x\in[0,1],\quad \int_0^{x_e} f(x) dx =0.99 n_e, \quad \int_0^1 f(x)dx =n_e.
\end{equation}
Function (\ref{fx}) is defined on the interval $[0,1]$ and it seems suitable to describe the fecundity as function of physiological age.
Different oviposition profiles, corresponding to different values of $x_*$ and $x_e$, are reported in Figure \ref{FigOvProf}. In Figure \ref{op02}, individuals entering in the adult stage become immediately reproductive and the peak of oviposition is in the first part of the adult physiological age. It represents a case with no pre-oviposition period. In the cases of fecundity represented in Figures \ref{op06} and \ref{op08}, adults start later in physiological age to produce eggs. The initial period in which they do not produce eggs can be seen as a pre-oviposition period. This way to model the fecundity rate is useful in case of existence of both pre-oviposition and reproductive adults having the same development function without introducing  a further stage of pre-oviposition adults. 
Fecundity profiles like the ones reported in Figures \ref{op02} and \ref{op06} also allow to include an eventual post-oviposition period for the adult stage. In the profile \ref{op08} the peak of the oviposition is strongly delayed, and it is suitable to empathize the effects of a late reproduction or a long pre-oviposition period.\\

It is also possible to consider a dependence of the fecundity function on temperature, as in \eqref{Fmod_xT} where $b(T)$ is the temperature profile. A possible expression can be given by the parabola \cite{gutierrez2012, gilioli2016}
\begin{equation}
b(T)=
\begin{cases}
1-\left(\dfrac{T-T_L-T_0}{T_0}\right)^2,& T_L\leq T\leq T_L+2T_0,\\
0,&\textnormal{otherwise},
\end{cases}
\label{bT}
\end{equation}
where $T_L$ and $T_L+2T_0$ are temperature thresholds for the egg production, determining the temperature range in which eggs are laid, while the optimal temperature is $T_L+T_0$.
Usually, the temperature interval of eggs production is enclosed in the temperature interval of positive development of adults $[T_m,T_M]$. 

In the numerical simulations, parameters values of the oviposition profile obtained with $n_e=100$ are reported in Table \ref{Tabf}, while for the temperature profile we choose
$$T_L=18.3\;  ^\circ C, \quad  T_0=6.5\;  ^\circ C.$$
\begin{table}[!h]
\centering
\begin{tabular}{cccc}
\toprule
	    & $x_*=0.2,\;x_e=0.6$   & $x_*=0.6,\;x_e=0.8$   & $x_*=0.8,\;x_e=0.95$ \\
\midrule
$\alpha$ & $2.86245\cdot 10^{4}$ & $2.3949016\cdot 10^{9}$ & $2.190076\cdot 10^{5}$ \\
$\beta$     & $1.8$                 & $13.9$               & $10.2$              \\
$\gamma$    & $7.2$                 & $9.27$              & $2.55$             \\
\bottomrule
\end{tabular}
\caption{Parameters of the oviposition profile for different values of $x_*$ and $x_e$ obtained for $n_e=100$.}
\label{Tabf}
\end{table}

\subsubsection{Mortality}
\begin{figure}%
\centering
  \subfigure[Eggs]{\includegraphics[width=0.35\textwidth]{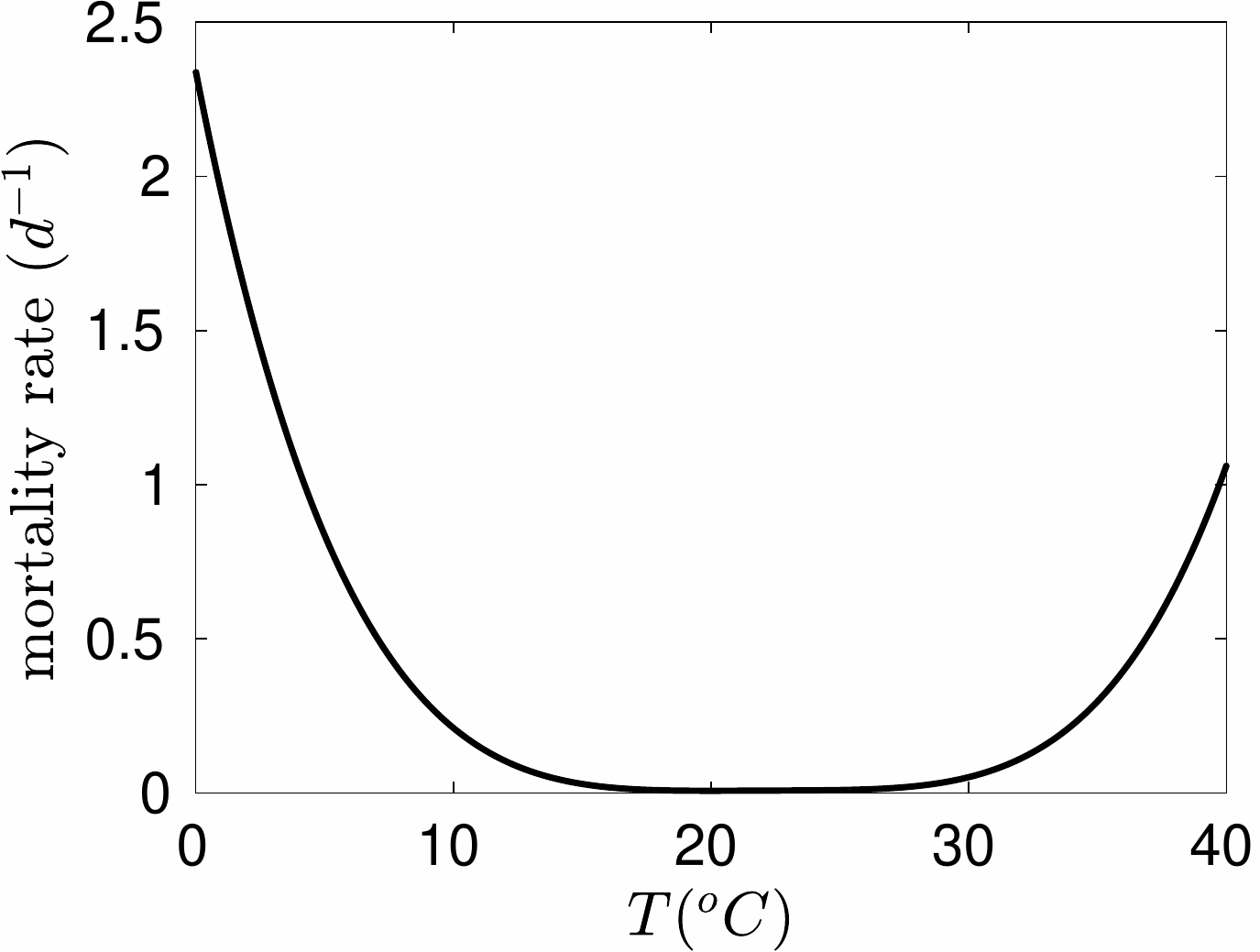}}
	\subfigure[Larvae]{\includegraphics[width=0.35\textwidth]{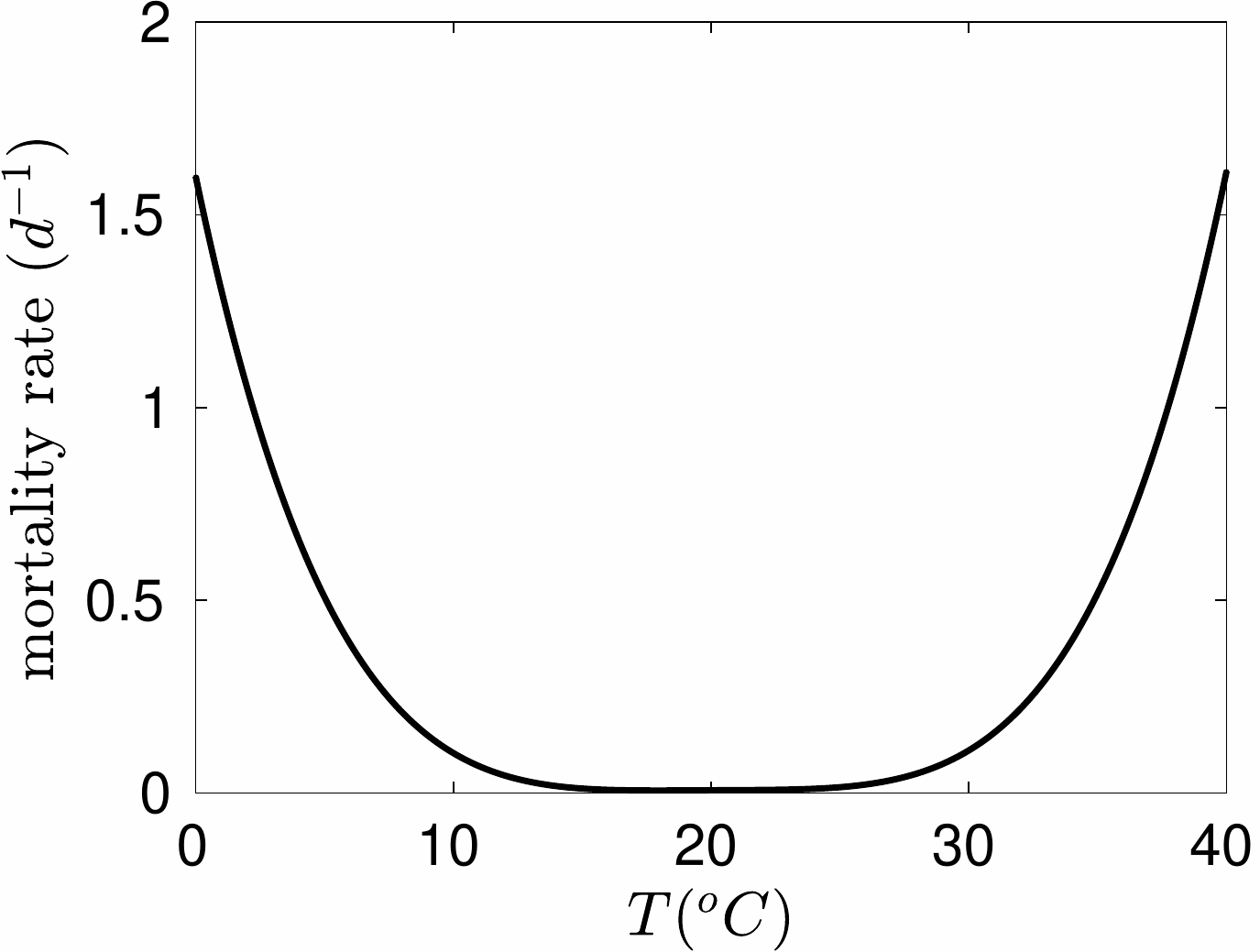}}
	\subfigure[Pupae]{\includegraphics[width=0.35\textwidth]{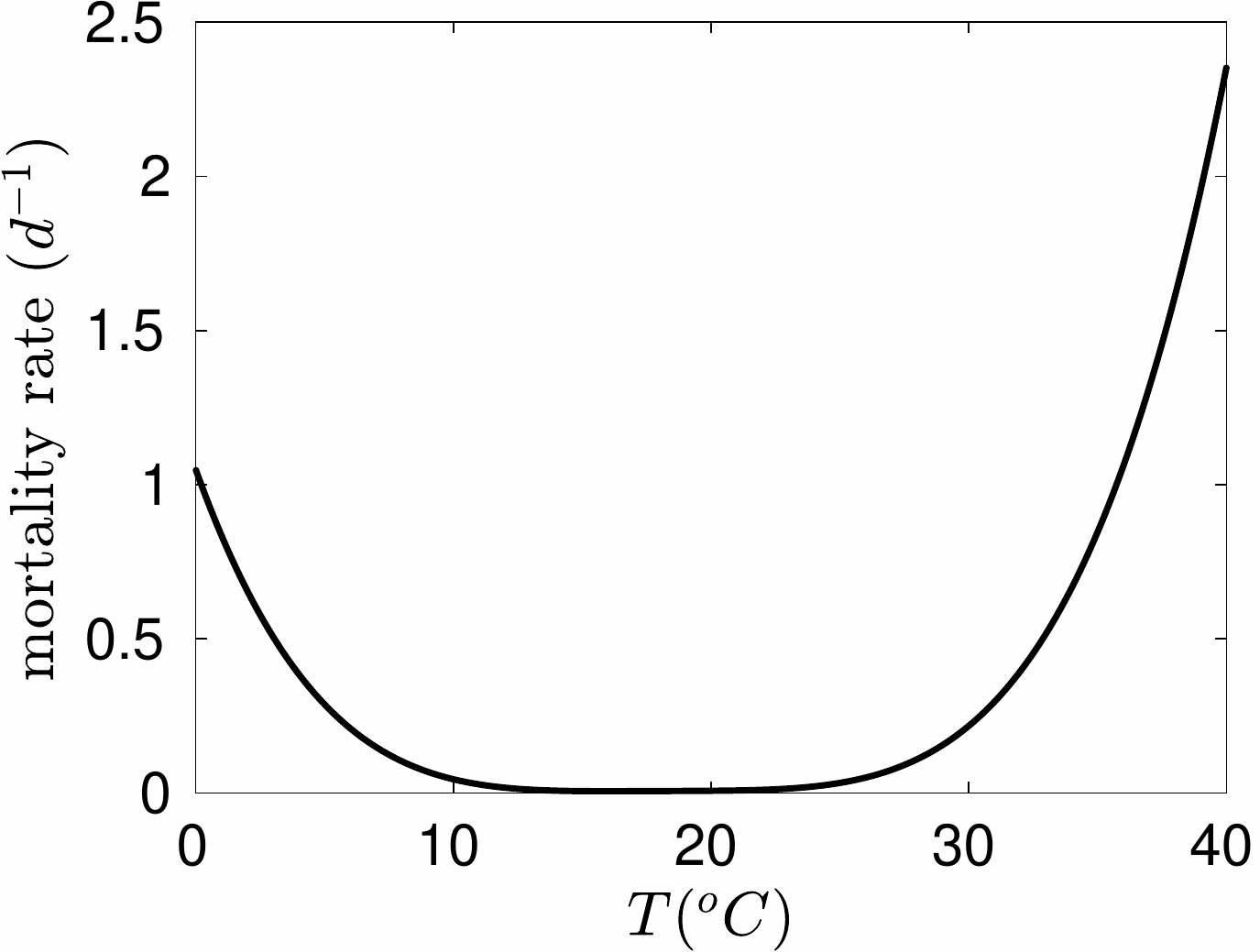}}
	\subfigure[Adults]{\includegraphics[width=0.35\textwidth]{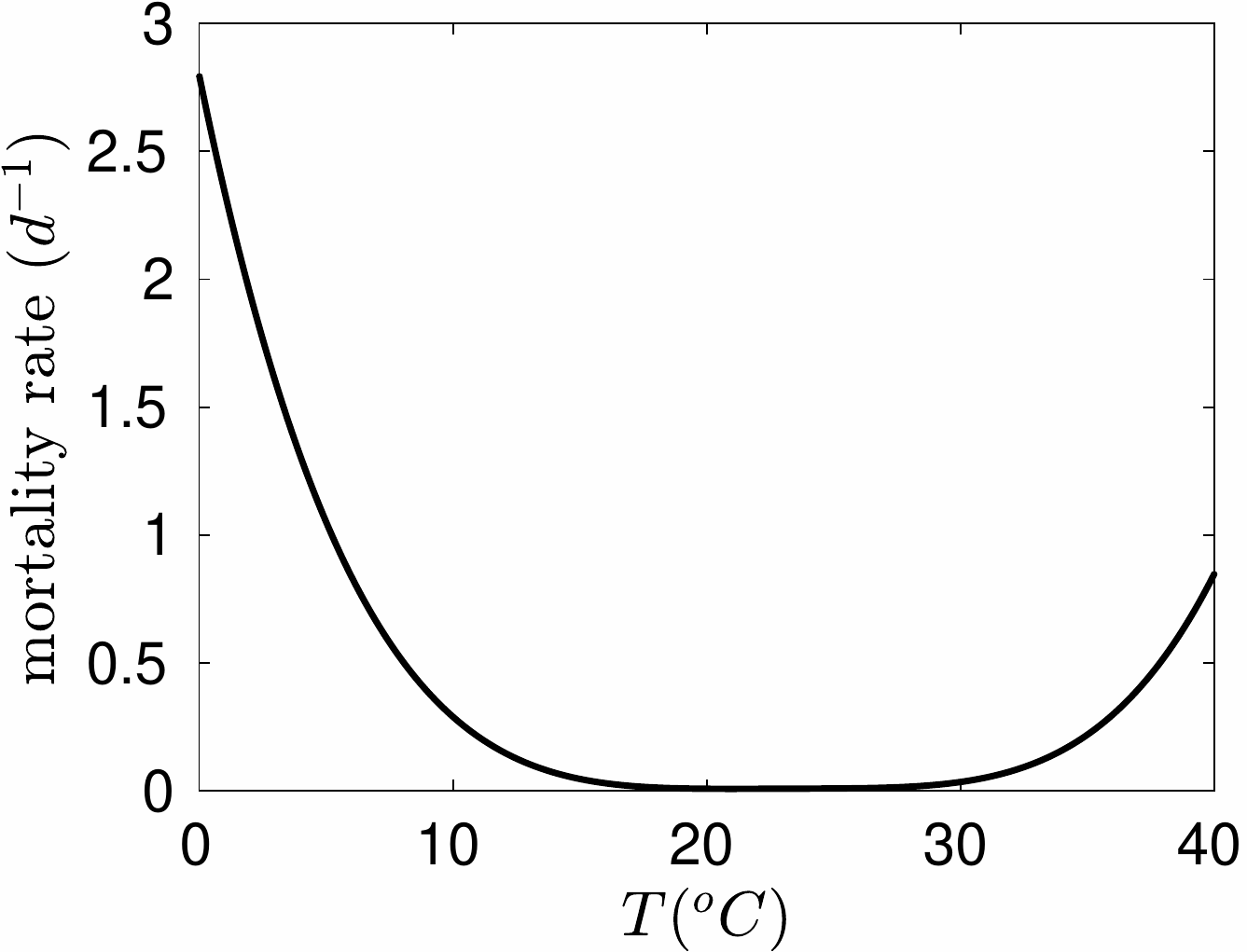}}
\caption{Mortality rate ($1/days$) as function of temperature ($^\circ C$) for the four stages.}
\label{FigM}
\end{figure}

The mortality rate is described by a bathtub shaped function \cite{wang2002}: this allows to have a small mortality rate for a favorable range of temperatures, and mortality rate that rapidly increases outside this interval.
The common expression chosen for the mortality rate for all the stages is 
\begin{equation}m(T)=\mu\left(b+cT+d(T_c-T)^4\right),\label{eq:mf}\end{equation}
where $\mu=0.005,\; b=0.00015,\; c=0.07,\; d=0.002$, while the parameter $T_c$ is considered different for each stage. The assumed values are reported in Table \ref{TabParM}, while the corresponding profiles 
are shown in Figure \ref{FigM}. It is worthwhile to note that in our simulations, the mortality has effect only after a fixed date (e.g., May $1^{st}$ in temperate regions), corresponding to a date in which the individuals have overcame the diapausing stage. 

\begin{table}[!h]
\centering
\begin{tabular}{ccccc}
\toprule
	    & $i=1$ & $i=2$ & $i=3$ & $i=4$ \\
\midrule
$T_c\; ( ^\circ C)$ & $22$ & $20$ & $18$ & $23$\\
\bottomrule
\end{tabular}
\caption{Parameter $T_c$ of the stage-specific mortality rate function in \eqref{eq:mf} for the four stages: 
 eggs ($i=1$), larvae ($i=2$), pupae ($i=3$) and adults ($i=4$).}
\label{TabParM}
\end{table}

\subsubsection{Initial conditions}
\begin{figure}%
\centering
\includegraphics[width=0.9\textwidth]{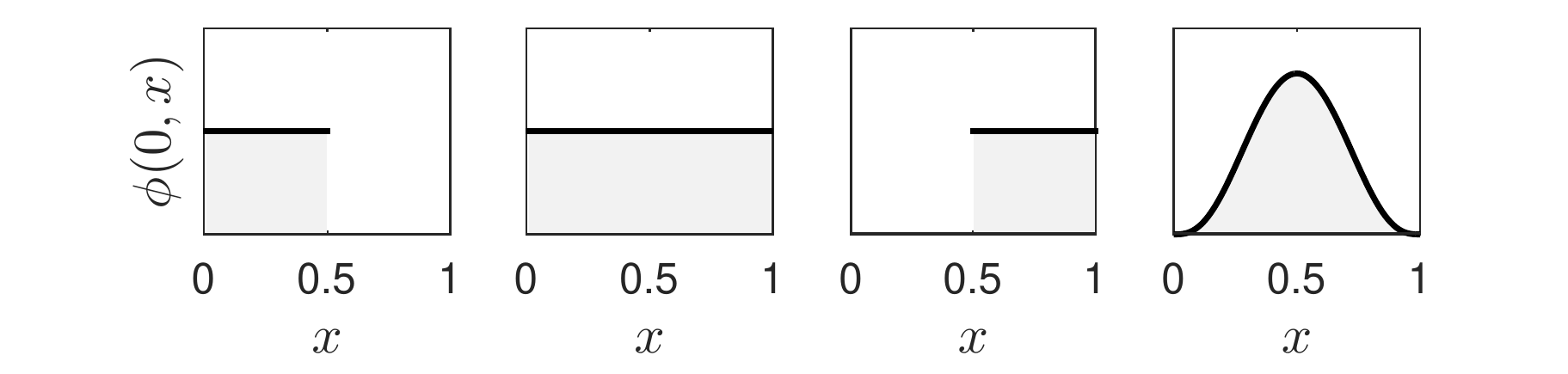}
\caption{Initial distributions with respect to the phisiological age of individuals in the overwintering stage.}
\label{FigInDistr}
\end{figure}

Different initial conditions (\ref{kolm4}) will be considered by defining the distribution of individuals along physiological age in each stage at the initial time $t_0$. If the initial time is January $1^{st}$, only the overwintering stage will have non zero value. 
In the following we will consider this case, starting the simulation from January $1^{st}$.
In Figure \ref{FigInDistr} we can see different choices for the initial distribution with respect to the physiological age. In the first case, individuals are equally distributed in the first half of the physiological age; this means that overwintering individuals enter diapause when young, then they will need a long period to complete the development in the overwintering stage. In the second case, individuals are equally distributed on the whole interval of the physiological age, modeling a situation in which individuals enter diapause at any age of the overwintering stage. In the third case, individuals are equally distributed in the second half of the physiological age; this means that overwintering individuals have already reached a certain percentage of maturation in the stage before enter diapause, or are next to the stage exit. Finally, we consider also a non uniform distribution on the whole interval $[0,1]$ obtained with a symmetric beta function. 

\subsection{Comparison of population dynamics under different models}\label{NumSim}
We decided to analyze separately the different models M1, M2, M3, to better understand the effects of variations in the initial condition and changes due to the introduction of mortality and fecundity. Combination of different models (that is, joint variation in mortality, fecundity and initial condition) are not considered here, but the effects on the dynamics can be easily guessed.

First of all, we investigate the role of the initial distribution, corresponding to model M1
for different intial conditions and comparing the obtained models with model M0 that has fecundity equal to adult development and initial condition of 100 pupae with physiological age zero. Then, we compare the three different formulations \eqref{Fclassico}, \eqref{Fmod_x} and \eqref{Fmod_xT} of the fecundity rate, corresponding to model M2, where the initial condition is set to 100 individuals equally distributed on the physiological age for the pupal stage, and equal to 0 for the other stages. In all cases, we do not consider the mortality effects, then $m^i(t)=0,\; i=1,\dots,4$. Finally, again starting from an initial condition of 100 pupae equally distributed on the physiological age, we consider the effect of the mortality analyzing model M3. 

The study is relative to a temperate climate region that require a diapause during the winter season. We suppose that the pupal stage is the overwintering stage and that the end of diapause is May $1^st$.

\begin{figure}%
\centering
\includegraphics[width=0.45\textwidth]{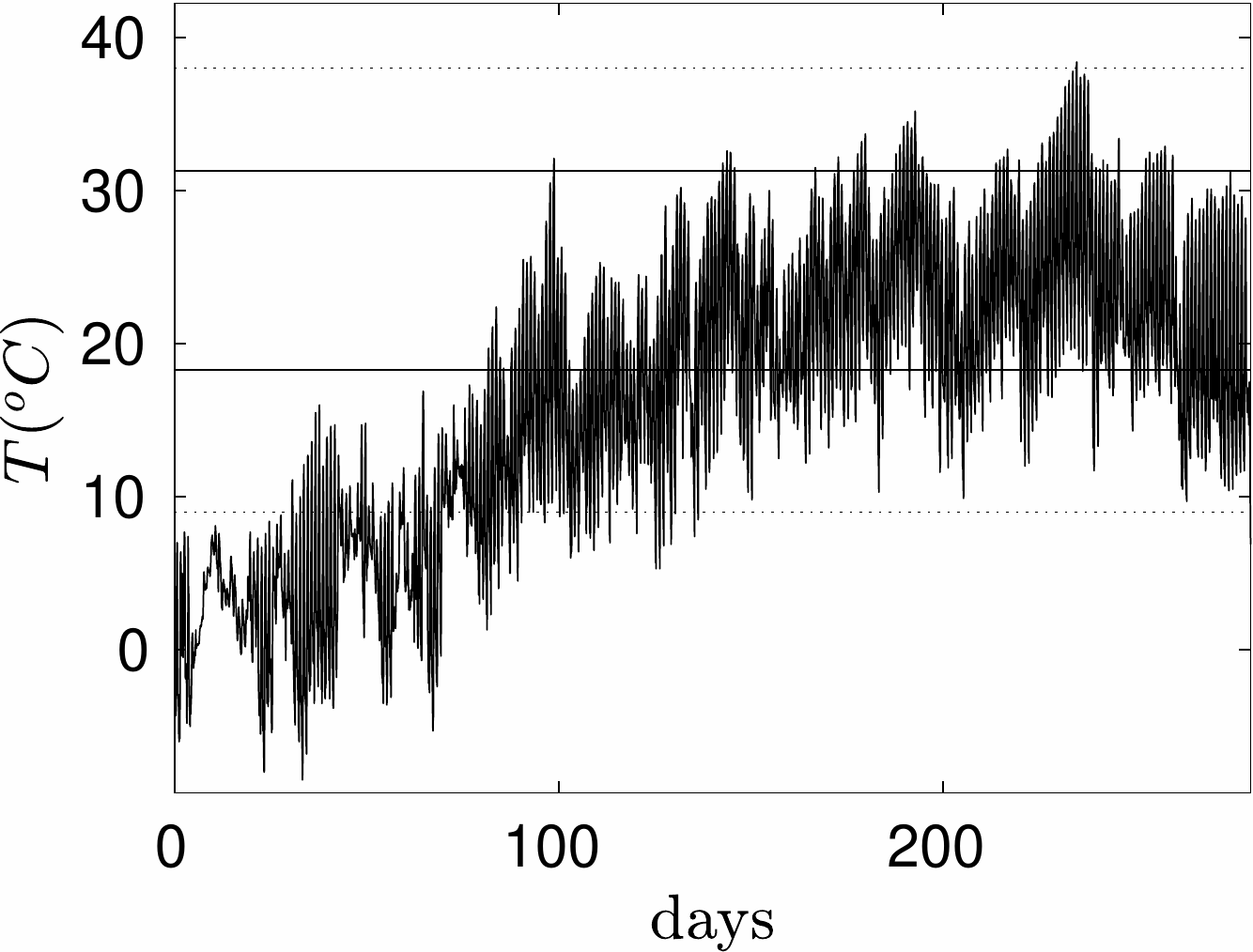}
\caption{Temperatures for the year 2011 of a weather station in Colognola ai Colli (North of Italy). Day 0 corresponds to January, $1^{st}$. Solid and dotted horizontal lines represent fecundity and adult development temperature thresholds, respectively.}
\label{T2011}
\end{figure}

For the numerical simulations we use a set of hourly temperatures from a weather station in Colognola ai Colli in the north of Italy of the year 2011 (Figure \ref{T2011}).

\subsubsection{Model M1}
\begin{figure}%
 \centering
 \includegraphics[width=1\textwidth]{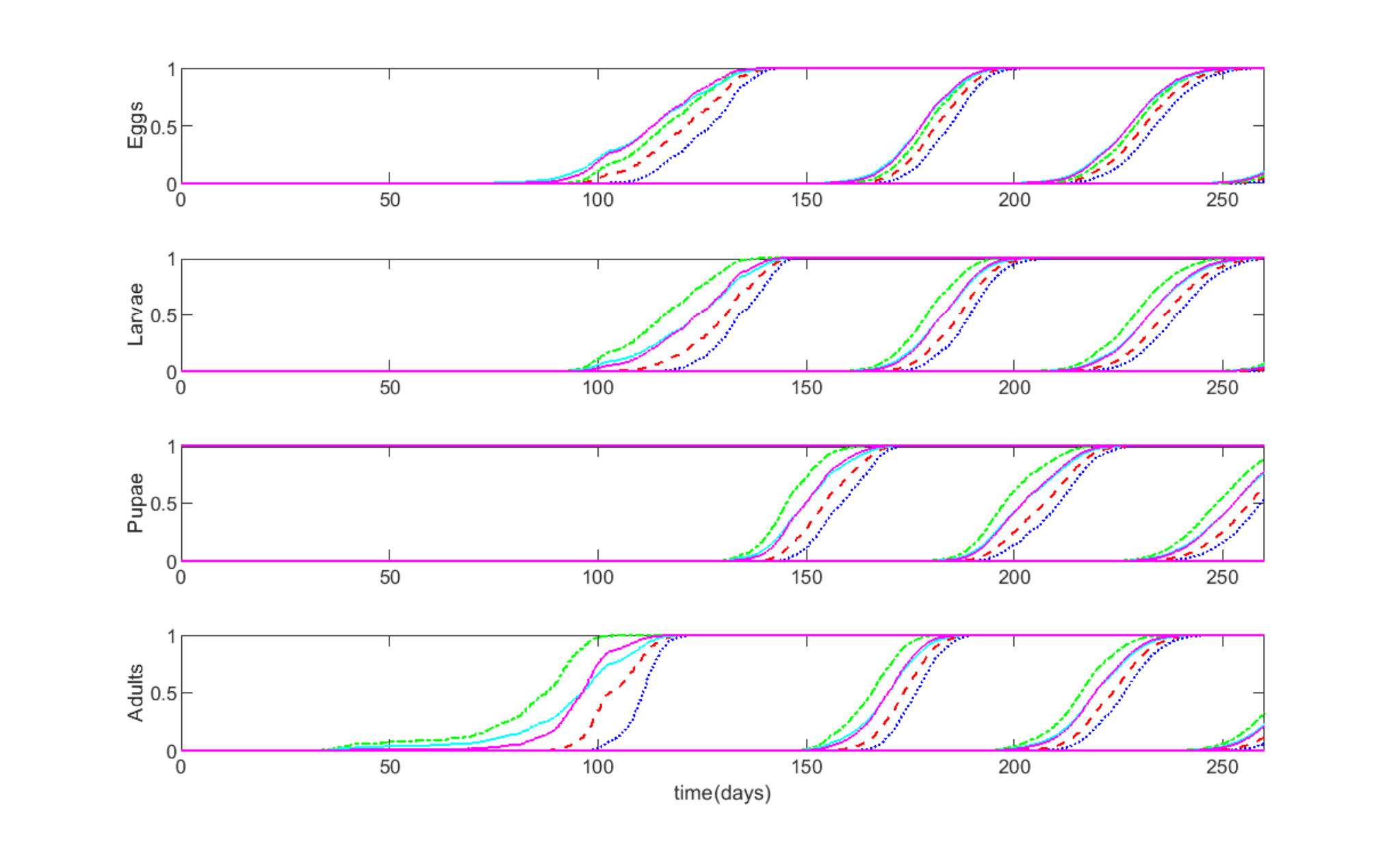}
 \caption{Cumulative percentage of individuals entering in the four stages (eggs, larvae, pupae, adults). Comparison among the model M0 with all the initial individuals having physiological age zero (blue dotted line) and model M1 for different initial distributions with respect to the physiological age of individuals in the initial stage. Green dashed-dotted line: uniformly distributed in the interval $[0.5,1]$. Light blue continuous line: uniformly distributed in $[0,1]$. Magenta continuous line: symmetric beta-distribution in $[0,1]$. Red dashed line: uniformly distributed in $[0,0.5]$. Day 0 corresponds to January, $1^{st}$.}
 \label{FigInCond}
 \end{figure}

The results of the numerical simulations obtained from different initial conditions are shown in Figure \ref{FigInCond} where the cumulative input flux entering in a stage, as a percentage of the total population, has been represented. Phenological dynamics for different distributions of the initial condition are compared with the dynamics obtained for model M0 where all the initial individuals have physiological age zero (blue dotted line). All the distributions represented in Figure \ref{FigInDistr} have been considered. 
Each curve is non decreasing (regression to a previous stage is not allowed), starts from 0\% and ends at 100\%, corresponding to the situation in which all the individuals have been already moved from the previous stage to the current. Different generations are evident in the figure. 
The most anticipated curve for larvae, pupae and adults (green dashed-dotted line), is relative to the initial condition uniformly distributed in the second half of the physiological age. The individuals exit from diapause when they have reached at least the 50\% of the development in the overwintering stage and thus the time required to change the stage is lesser than in the other cases. The next two curves are relative to the distribution of the initial stage over the whole physiological age interval. The adult dynamics obtained with the beta distribution of the initial condition (magenta continuous line) crosses the dynamics relative to the uniform distribution of the initial condition over the interval [0,1] (light blue continuous line).  More delay is observed for the curve with a uniform distribution of the initial condition over the first half of the physiological age (red dashed line) because the individuals exit from diapause when young and thus more time is required to change stage with respect to the distributions of the initial condition previously considered. The most delayed curve is, obviously, the one with initial condition of individuals of age zero which require more time to complete the development in the overwintering stage.

The larger gap among the different curves is in the first adult generation, while the gap tends to decrease in the next generations of each stage. 
Starting from the pupal stage, the effects of different initial conditions are more evident on the flux in the next stage, that is on the first fly of the adult stage after the winter season. The effects decrease for the first egg and larval generations, then the gap among different curves remain approximately constant for all the generations of each stage.
Quantitatively, for the adult stage, there is a gap of approximately 24 days from model M1 with initial condition uniformly distributed over the interval [0.5,1] of the physiological age and model M0. For the egg stage the gap is of approximately 14 days from model M1 with initial condition beta distributed over [0,1] and model M0.
Considering the curves obtained for an initial condition uniformly distributed over [0,1] (light blue continuous lines) and an initial condition beta distributed over [0,1] (magenta continuous line), we notice that, except for the adult first generation, the outcomes are similar. Then, there are no considerable differences in the dynamics with respect to a uniform or a symmetric non uniform distribution over the whole physiological age interval. The only difference is the slope of the first adult generation.

\subsubsection{Model M2}

\begin{figure}%
\centering
  \subfigure[]{\includegraphics[width=0.85\textwidth]{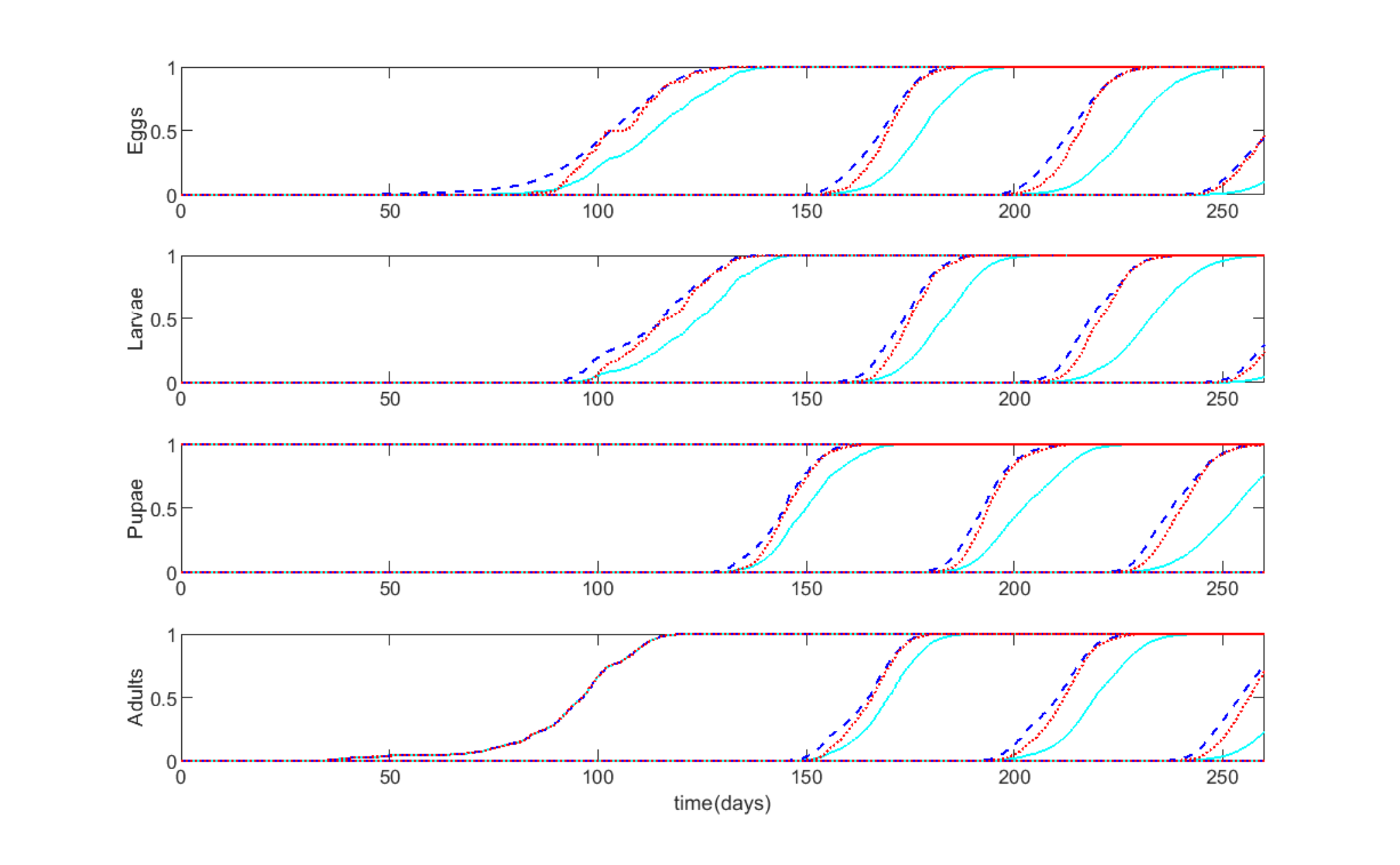}}\\
	\subfigure[]{\includegraphics[width=0.85\textwidth]{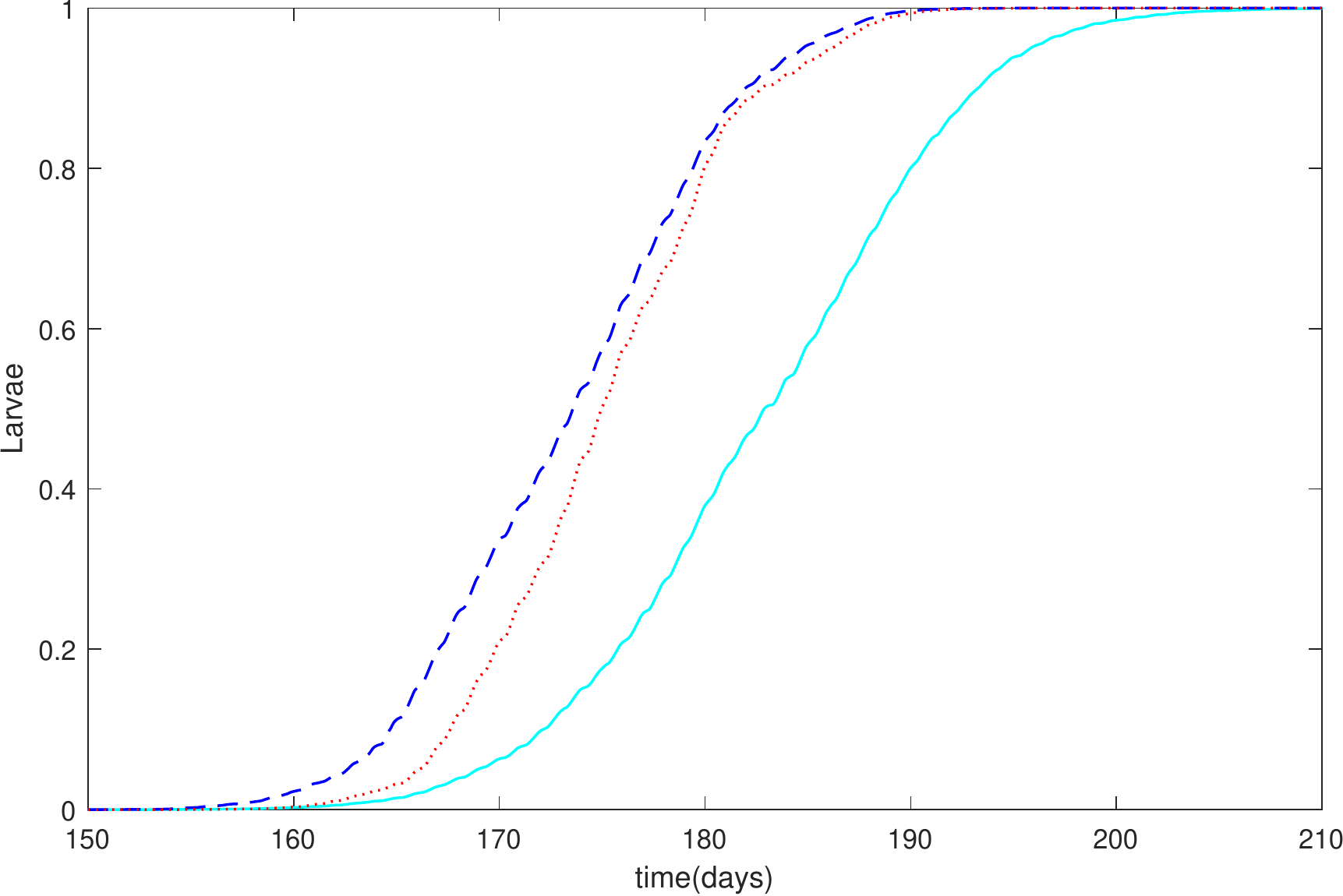}}
\caption{Cumulative percentage of individuals entering in the four stages (eggs, larvae, pupae, adults). Light blue continuous line: model \eqref{Fclassico}, fecundity equal to adult development. Blue dashed line: model \eqref{Fmod_x}, fecundity dependent on physiological age with the profile \eqref{op02}. Red dotted line: model \eqref{Fmod_xT}, fecundity dependent on temperature and on physiological age (with profile \eqref{op02}). Temperatures of a weather station in the North of Italy for the year 2011. Day 0 corresponds to January, $1^{st}$.}
\label{Fig02}
\end{figure}

\begin{figure}%
\centering
    \subfigure[]{\includegraphics[width=0.85\textwidth]{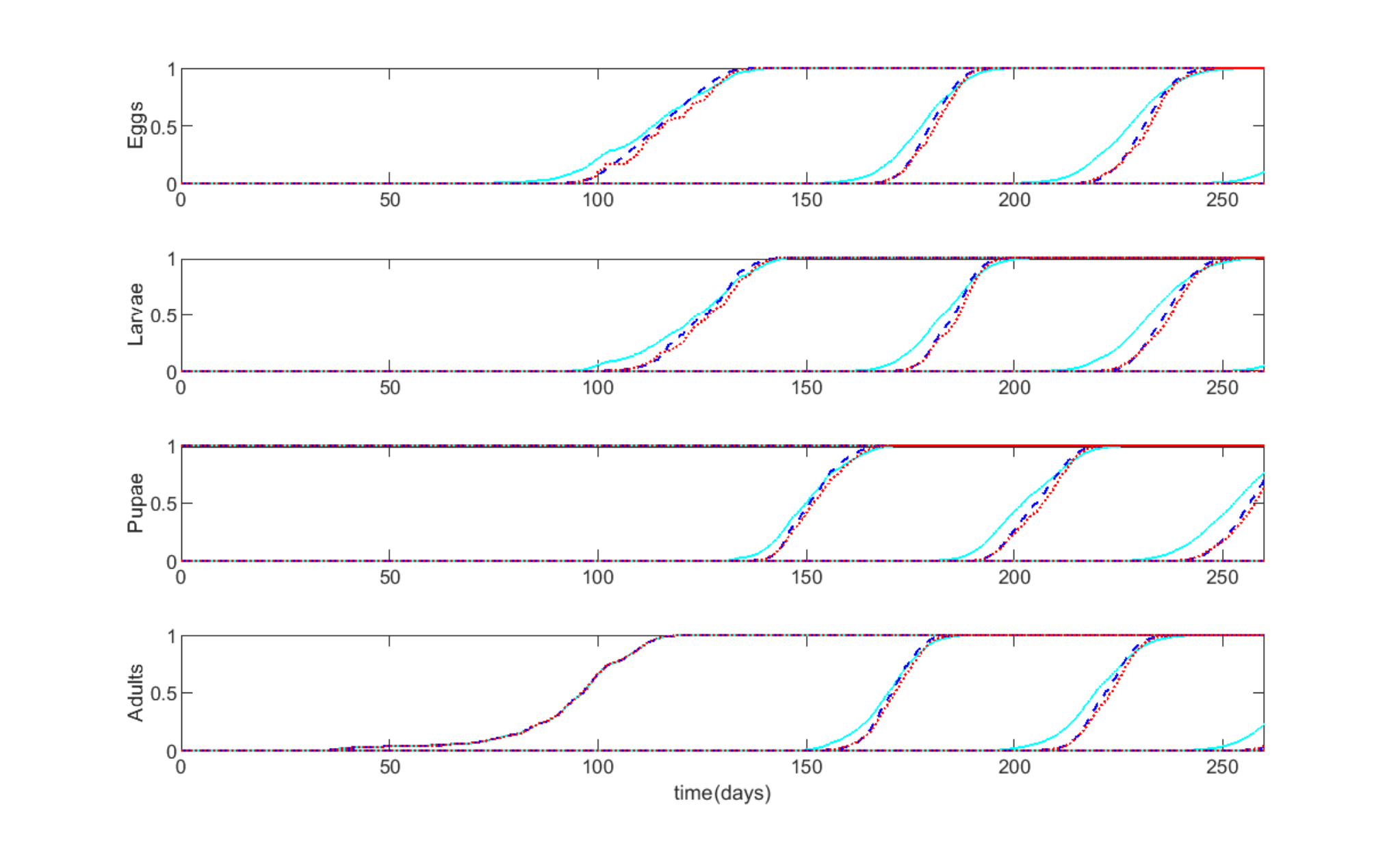}}\\
	\subfigure[]{\includegraphics[width=0.85\textwidth]{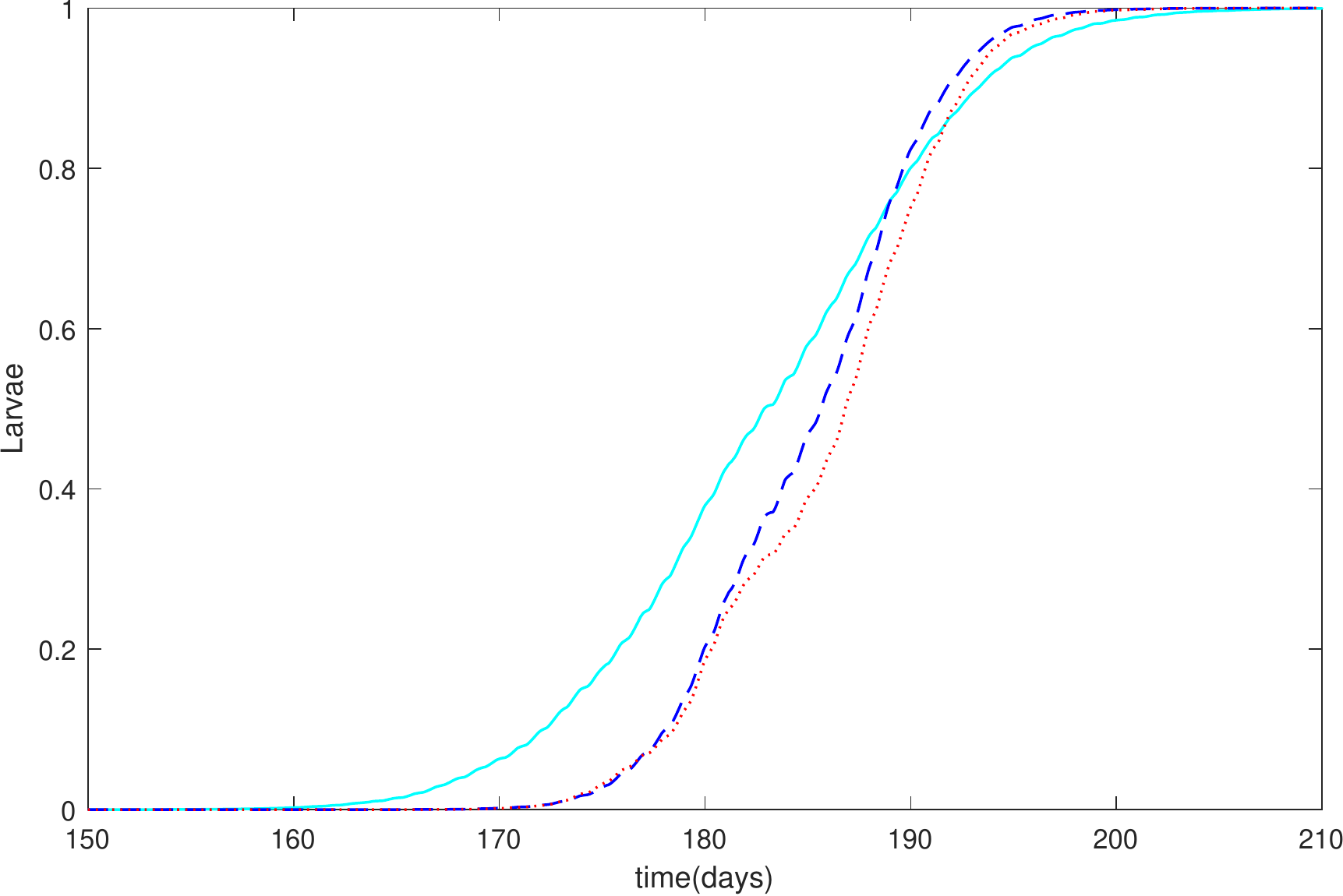}}
\caption{Cumulative percentage of individuals entering in the four stages (eggs, larvae, pupae, adults). Light blue continuous line: model (\ref{Fclassico}), fecundity equal to adult development. Blue dashed line: model (\ref{Fmod_x}), fecundity dependent on physiological age with the profile (\ref{op06}). Red dotted line: model (\ref{Fmod_xT}), fecundity dependent on temperature and on physiological age (with profile (\ref{op06})). Temperatures of a weather station in the North of Italy for the year 2011. Day 0 corresponds to January, $1^{st}$.}
\label{Fig06}
\end{figure}

\begin{figure}%
\centering
  \subfigure[]{\includegraphics[width=0.85\textwidth]{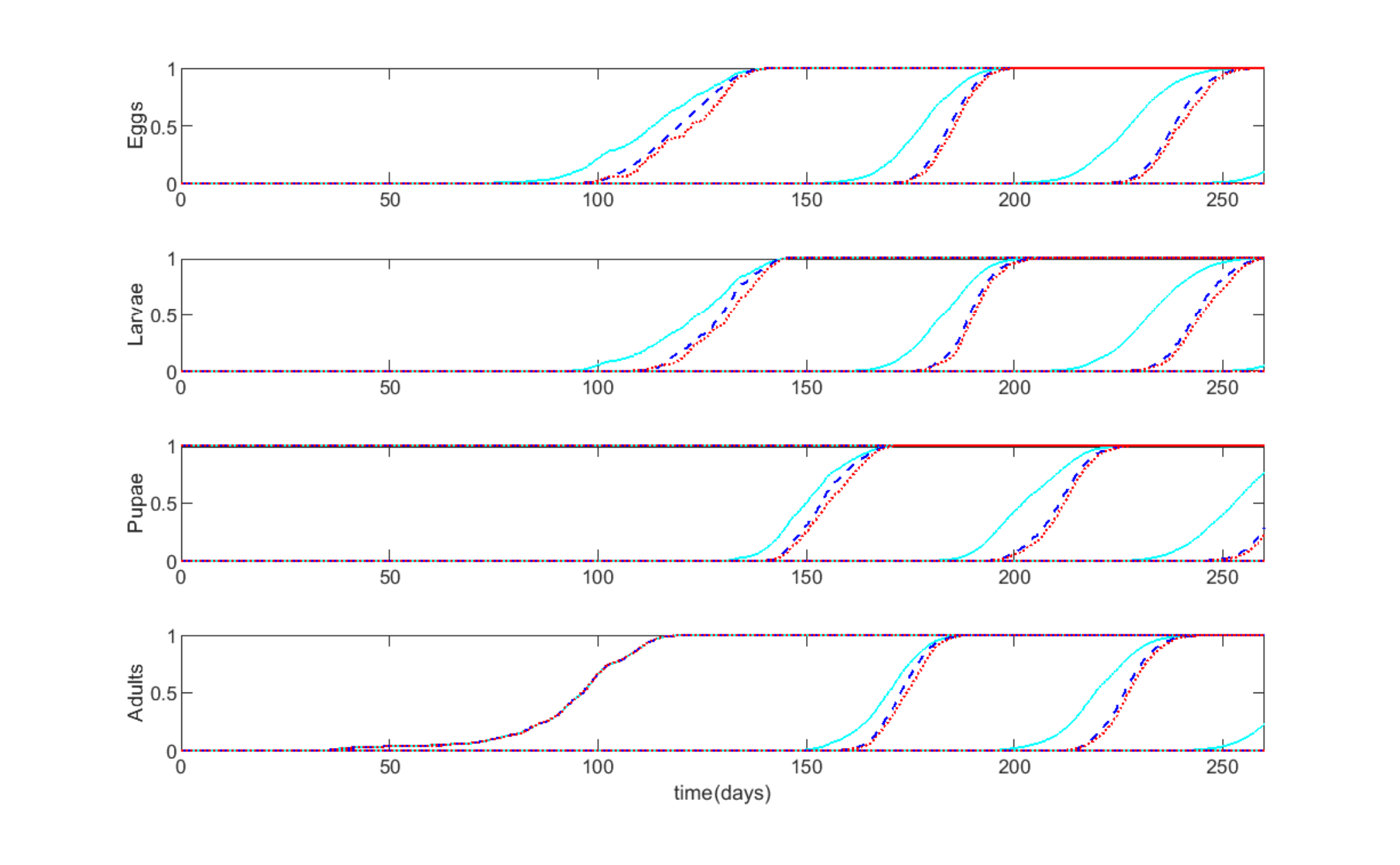}}\\
	\subfigure[]{\includegraphics[width=0.85\textwidth]{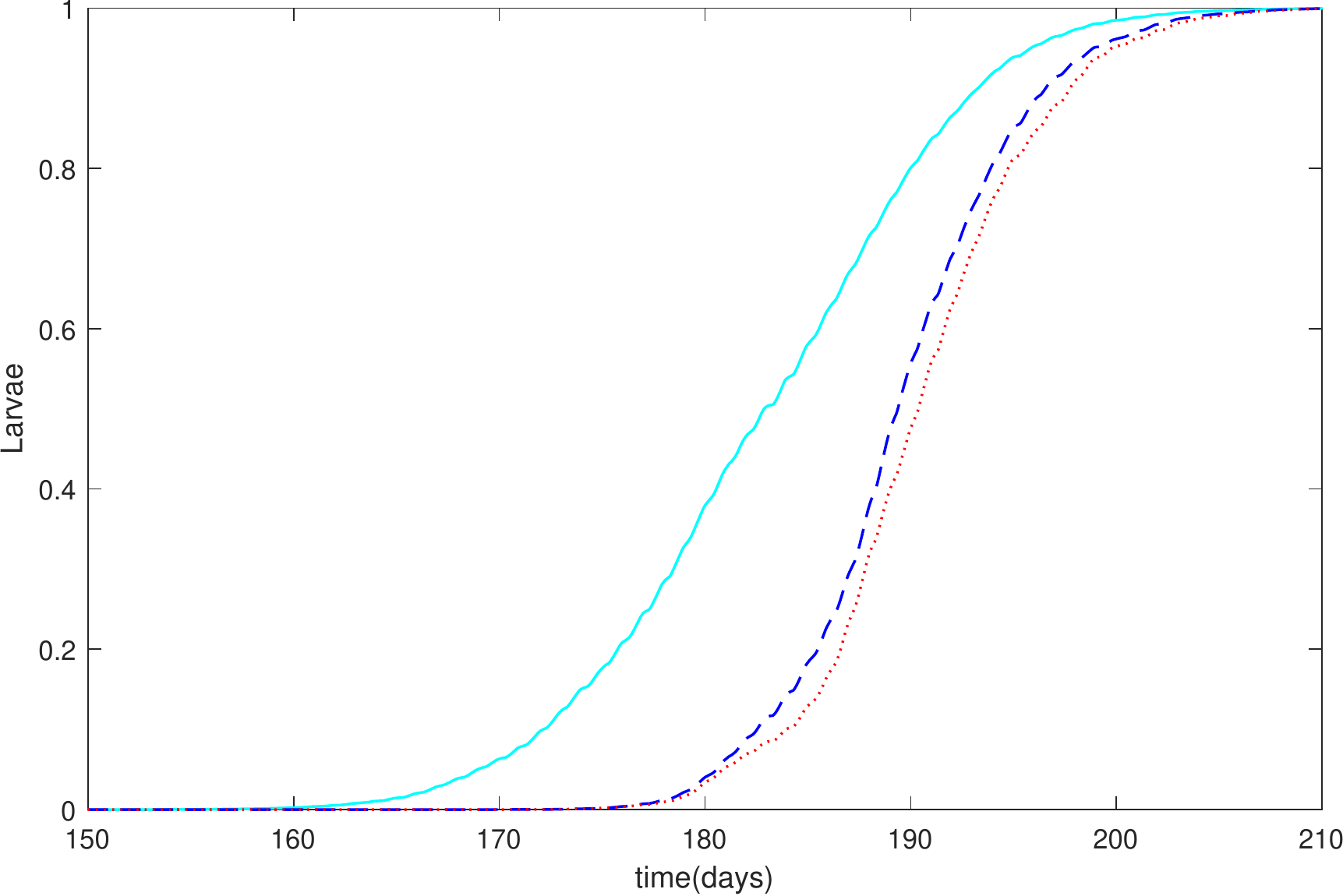}}
\caption{Cumulative percentage of individuals entering in the four stages (eggs, larvae, pupae, adults). Light blue continuous line: model (\ref{Fclassico}), fecundity equal to adult development. Blue dashed line: model (\ref{Fmod_x}), fecundity dependent on physiological age with the profile (\ref{op08}). Red dotted line: model (\ref{Fmod_xT}), fecundity dependent on temperature and on physiological age (with profile (\ref{op08})). Temperatures of a weather station in the North of Italy for the year 2011. Day 0 corresponds to January, $1^{st}$.}
\label{Fig08}
\end{figure}

In Figures \ref{Fig02}--\ref{Fig08}, starting from an initial condition of pupae uniformly distributed over the interval [0,1], we represent the normalized incoming flux in each stage  for various fecundity profiles (equations \eqref{Fclassico}, \eqref{Fmod_x} and \eqref{Fmod_xT}). Different generations are evident in the graphs. To better show the gap among different curves, we enlarge, as an example, the second generation of the larval stage. 

We note that the first adult generation is equal for each fecundity formulation, while starting from the first eggs generation there are differences in the outcomes. In detail, with the oviposition profile \eqref{op02}, the dynamics of the model \eqref{Fmod_x} are in advance with respect to the ones of model \eqref{Fclassico} (Figure \ref{Fig02}), and the gap is up to approximately 10 days on the second larval generation. With the oviposition profile \eqref{op06} (Figure \ref{Fig06}) model \eqref{Fclassico} is in advance in the first part of the dynamics and delayed in the second part but the gap is, in general, smaller than in the previous case. In case of oviposition profile \eqref{op08}, the dynamics of the model \eqref{Fmod_x} is delayed with respect to that of model \eqref{Fclassico} (Figure \ref{Fig08}), and the delay is up to approximately 10 days in the first part of the dynamics for the second larval generation.

Independently from the oviposition profile, the population dynamics obtained with a temperature dependent fecundity (continuous line) has a little delay with respect to the dynamics of the population obtained with a fecundity dependent only on physiological age (dashed line). 

\subsubsection{Model M3}
\begin{figure}%
\centering
    \subfigure[]{\includegraphics[width=0.85\textwidth]{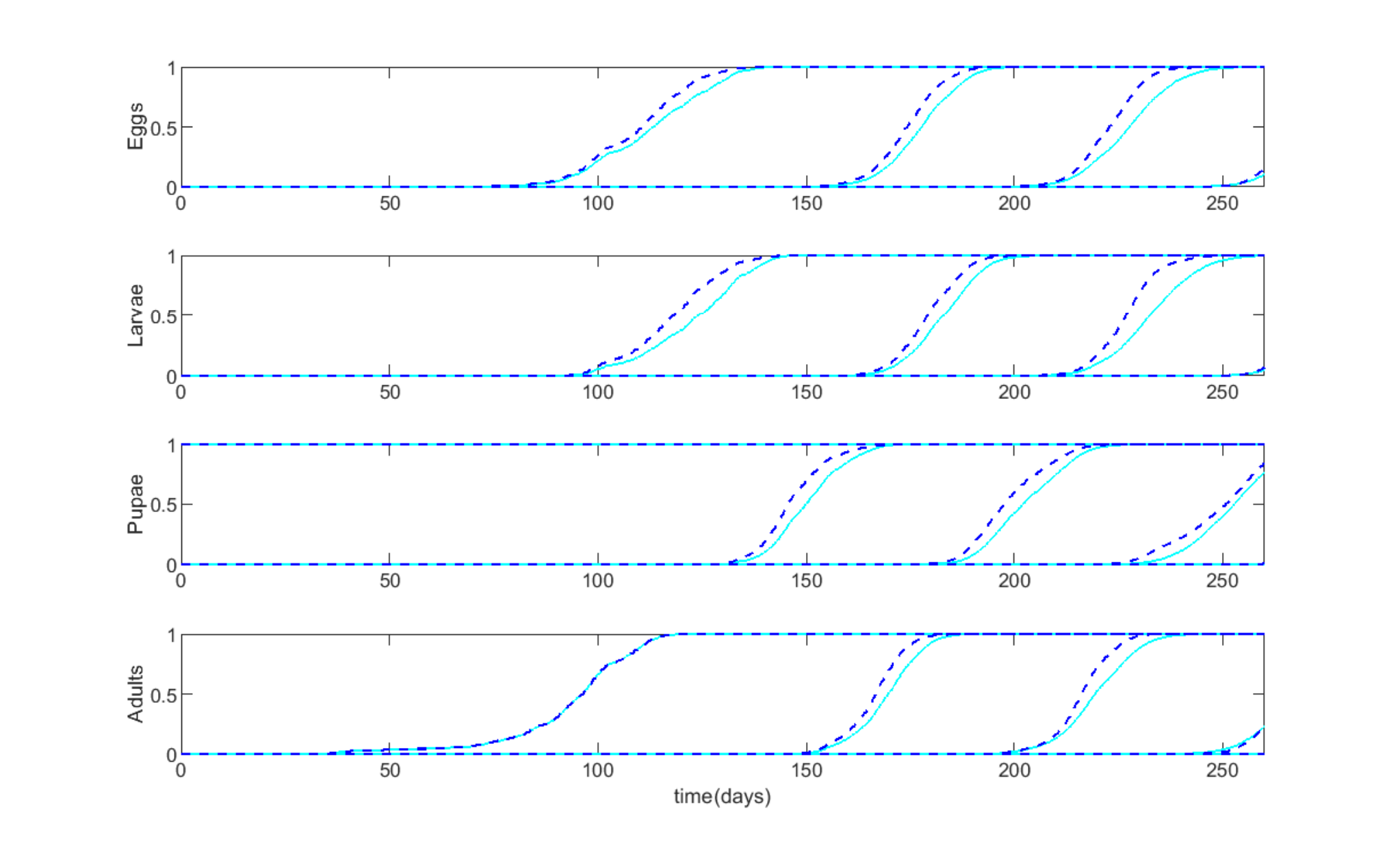}}\\
	\subfigure[]{\includegraphics[width=0.85\textwidth]{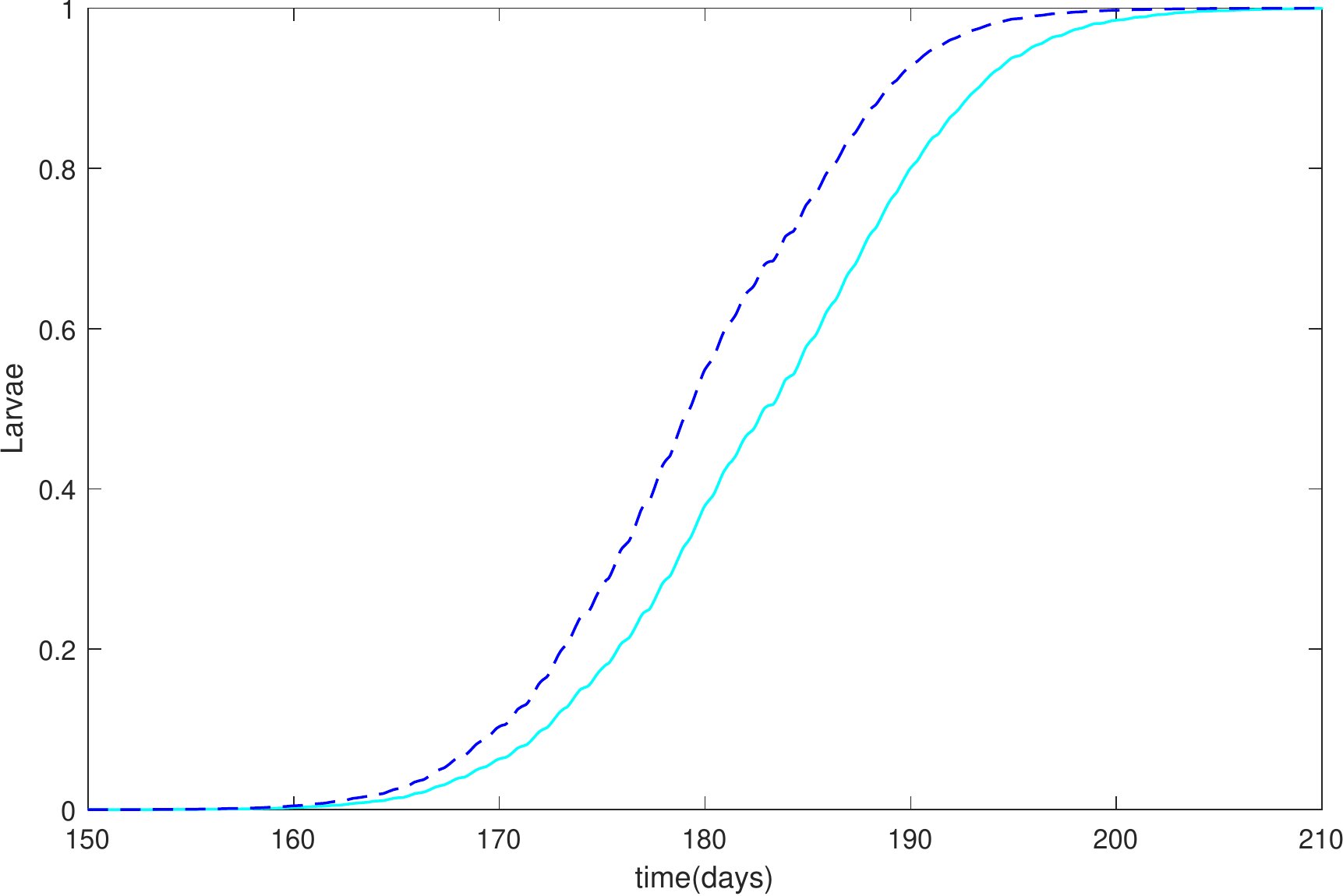}}
\caption{Cumulative percentage of individuals entering in the four stages (eggs, larvae, pupae, adults) with model \eqref{Fmod_xT}, fecundity dependent on temperature and on physiological age (with profile shown in Figure \ref{op02}). Blue dashed line: model with mortality. Light blue continuous line: model without mortality. Temperatures of a weather station in the North of Italy for the year 2011. Day 0 corresponds to January, $1^{st}$.}
\label{Fig02_M}
\end{figure}

In this section, we introduce mortality, in addition to fecundity dependent on temperature and physiological age with profile \ref{op02}. Starting from an initial condition of 100 pupae uniformly distributed over the physiological age interval $[0,1]$ on May $1^{st}$, the dynamics are in advance with respect to the dynamics obtained with the fecundity equal to adult development of model M1 (Figure \ref{Fig02_M}) and the gap is of the order of some days for the second generation of the larval stage.
This is justified by the fact that if the individuals cannot die, more time is required before all individuals change the stage.



\section{Application to a case study: the phenology of the codling moth}
\label{cydia}

To further explore the consequences of different phenological model formulation we consider a case study of a key pest in apple orchards, the codling moth \emph{Cydia pomonella} (lepidoptera: Tortricidae) \cite{CABI}.

As stated in \cite{aghdam2009b}, phenological models have been already used to predict many phenological events related to the development of the different stages of the codling moth and to regulate pesticide treatments to control the codling moth.


We show how the use of a phenological model including the effects of fecundity can be useful to better describe and fit the real field data than the phenological model based only on developmental rates. 
An unpublished dataset of population dynamics limited to the adult stage collected in an apple orchard located in Gambellara (Ravenna), a flat region of Italy, during the year 2017 is considered. The experimental field was not treated with insecticides to avoid effect on pest population phenology and dynamics. A set of hourly temperatures from the closest weather station of San Pietro in Vincoli (Ravenna), approximately 5 Km far from Gambellara, is used.

We consider in our model formulation, four stages ($s=4$); the initial population is placed in the larval stage which is the overwintering stage. As development rate functions, for all the stages, we choose Lactin functions \cite{lactin1995}:
\begin{equation}
v(T)=\begin{cases}
\exp(\rho T) - \exp\left(\rho T_{max}-\dfrac{T_{max}-T}{\Delta}\right), & T\leq T_{max},\\
0, & T>T_{max}.
\end{cases}
\label{eq:lactin}
\end{equation}
Parameters of the development rate functions are estimated by means of a least square method using the data obtained in lab condition at individual level in \cite{ranjbar2009,aghdam2009} and estimated values are reported in Table \ref{TabDevCy}.

\begin{table}[!h]
\centering
\begin{tabular}{ccccc}
\toprule
	    & $i=1$ & $i=2$ & $i=3$ & $i=4$ \\
\midrule
$\rho$ & $0.173$ & $0.151$ & $0.160$ & $0.119$\\
$T_{max}\; ( ^\circ C)$ & $36.759$ & $37.094$ & $37.763$ & $42.905$               \\
$\Delta$ & $5.771$ & $6.628$ & $6.238$ & $8.353$              \\
\bottomrule
\end{tabular}
\caption{Parameters of the stage-specific development rate function in \eqref{eq:lactin} for the four stages of \emph{Cydia pomonella}:
 eggs ($i=1$), larvae ($i=2$), pupae ($i=3$) and adults ($i=4$).}
\label{TabDevCy}
\end{table}

The fecundity function depends on both the physiological age and the temperature. The parameters in formulas \eqref{fx} and \eqref{bT} are estimated from data in \cite{aghdam2009}. In particular, we chose the oviposition profile corresponding to Figure \ref{op06}, that allows to take into account a pre-oviposition period, as indicated by the experimental data. We set $\alpha=1$ and we fit the parameters $\beta, \; \gamma$ in \eqref{fx} by imposing the constraints in \eqref{vincoli_f} with $n_e=1$. Then we used the data on the mean number of eggs laid by a female at different constant temperatures $(T_k,n_e^k), \; k=1,\dots,N,$ (\cite{aghdam2009} and reported in Table \ref{TabEggsCy}) to estimate parameters $T_L,\; T_0$ in \eqref{bT}. Finally, we find $\alpha$ in \eqref{fx} that solve the problem
$$ \min_{\alpha>0} \sum_{i=k}^{N}{\left(\int_0^1f(x)b(T_k)dx-n_e^k\right)^2}.$$

\begin{table}[!h]
\centering
\begin{tabular}{ccccc}
\toprule
$T_k\; ( ^\circ C)$ & $20$ & $25$ & $27$ & $30$\\
$n_e^k$ & $48.846$ & $89.250$ & $66.316$ & $20.182$\\
\bottomrule
\end{tabular}
\caption{Mean number of eggs laid by a female at different constant temperatures $T_k$ obtained in lab condition at individual level \cite[pag. 235]{aghdam2009}.}
\label{TabEggsCy}
\end{table}

We obtain:
$$\alpha=2.064\cdot 10^{9},\; \beta=13.9,\; \gamma=9.267,\quad  T_L=17.755\; ^\circ C,\; T_0=6.499\; ^\circ C.$$

The effects of mortality are here neglected, because reliable data for estimating mortality rate function for all the stages are not available. 

\begin{figure}%
\centering
    \includegraphics[width=1\textwidth,trim={2.4cm 0 2.7cm 0},clip]{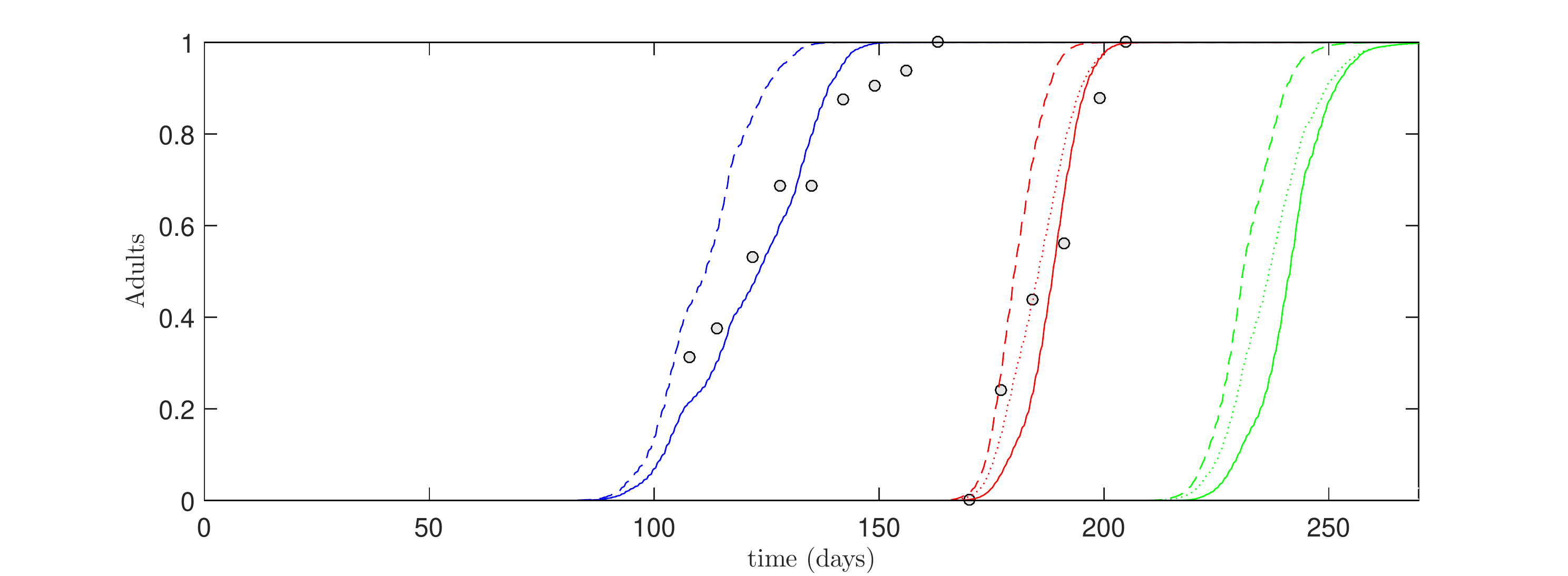}
\caption{Cumulative percentage of individuals entering in the adult stage for the pest \emph{Cydia pomonella} . Dashed line: model M1 with fecundity \eqref{Fclassico} equal to adult development, and uniform initial distribution of overwintering individuals on the second half of the physiological age. Dotted line: model M1 with fecundity \eqref{Fclassico} and uniform initial distribution of overwintering individuals on the whole interval of the physiological age. Solid line: model M1 with fecundity \eqref{Fmod_xT} dependent on temperature and on physiological age (with profile \ref{op06}), and uniform initial distribution of overwintering individuals on the whole interval of the physiological age. Temperatures of a weather station in San Pietro in Vincoli (Italy) for the year 2017. Day 0 corresponds to January, $1^{st}$.}
\label{FigAdCy}
\end{figure}

The results of the numerical simulations are shown in Figure \ref{FigAdCy} and compared with field data (marked with grey points). 

As reported in \cite{khani2007} we set the initial conditions such that the codling moth overcomes the winter as a full-grown diapausing larva. We start from model M1, with fecundity\eqref{Fclassico}, and initial population composed by 100 individuals uniformly distributed in the second half of the physiological age of the overwintering larval stage. The simulated dynamics (dashed lines in Figure \ref{FigAdCy}) present an advance for the first and second generations. 
Then, to delay the generations, we modify the initial condition by considering 100 larvae uniformly distributed in all the interval of the physiological age. The simulated dynamics (dotted lines) show a better fit for both the first and second generations. Then we introduce a fecundity dependent on temperature and physiological age (equation \eqref{Fmod_xT}). The dependence on temperature is of the form \eqref{bT}, while the dependence on physiological age is of the type represented in Figure \ref{op06} to take into account the pre-oviposition period of such species \cite{ranjbar2009,aghdam2009}. As initial condition we maintain 100 larvae uniformly distributed over the whole physiological age interval that allows to well fit the first generations. The simulated dynamics (solid lines) move forward with respect to the previous case. 
Simulations show that a change in the distribution of the initial population over physiological age produces a considerable shift in the adult dynamics of \emph{Cydia pomonella}. Moreover, the introduction of the fecundity further delays of some days the adults.

\section{Concluding remarks}\label{Concl}

In this paper we analyze various formulations of phenological models describing time variation of the stage structure of a poikilotherm population. The aim is to compare model performance in relation to assumption and process on which the model is built. In particular, starting from the basic formulation in which only temperature-development rate function is considered and all the individuals of the initial population have physiological age zero,  we investigated the effect of the introduction of fecundity and mortality rate functions, as well as the change in the initial conditions in terms of age distribution of the individuals entering in the overwintering stage.
The reference phenological model for this analysis is obtained as a simplification of the demographic model for stage structured populations presented in \cite{BuffoniPasquali2007}. The model, described by a system of partial differential equations, is driven by the ambient temperature and describes the time variation of the percentage of individuals in each stage, thus it does not require any knowledge of the number of individuals in each stage at a certain point in time to initialize the model, as it is needed for the demographic model.
The basic formulation of the phenological model takes into account the development rate functions and a fecundity one to one (each adult generates an egg) with a fecundity rate equal to the adult development rate. 
The absence of mortality guarantees that the number of individuals remains constant over time.
Then in the model are introduced more complex formulations  of the fecundity rate function and a mortality rate function is added. The mortality is applied only after the overwintering termination.

We observe that the introduction of the fecundity as function of the adult physiological age and temperature can anticipate or postpone the dynamics depending on the oviposition profile with respect to the physiological age (Figure \ref{FigOvProf}). The profile of the fecundity rate function is an important biological trait of a species and in many cases is known from rearing and lab observation of the adult females. The general pattern emerging from our numerical experiments is that the dynamics are anticipated with respect to those obtained for model M0 when considering a fecundity profile with peak corresponding to a young physiological age and delayed when the peak corresponds to a older physiological age. If the peak is located in the middle of the physiological age interval, the dynamics are in advance in the first part and then delayed in the second part. If the fecundity depends also on temperature a little delay in the dynamics is observed with respect to the dynamics obtained considering a fecundity dependent only on physiological age.

The proper definition of a fecundity profile has also to account for the existence and duration of  a pre- and/or a post-oviposition period. Describing these two periods and the fertile period by means of a single distribution  allows to reduce the number of stages of the structured population (reducing the number of differential equations). The introduction of the mortality generally anticipates the dynamics since less individuals require less time to leave the current stage than the whole population. Unfortunately, experimental or field data on mortality rates are not often available. In case of unavailability, the mortality function cannot be calibrated and sensibly introduced in the model.

The consideration of the age distribution of the initial population at January $1^{st}$, which corresponds to the age distribution of the overwintering stage, leads to important changes in the phenology of the simulated population. If the initial population density is concentrated towards the beginning of the overwintering stage, more time is required to develop and we observe a delayed pattern in the phenology, while if the initial population is composed by older individuals towards the end of the stage, the observed phenology is anticipated (Figure \ref{FigInCond}). For multi-voltine species the effect of the age distribution at the beginning of the overwintering is evident in the phenological pattern in the first generation appearing after the overwintering period. After the first generation this effect is negligible. 

The analysis carried out shows the relevance of introducing supplementary factors in a basic phenological model. When elements of biological realism like fecundity, mortality and initial age distribution are introduced the model output changes. This gives rise to the issue about on when and how to consider the additional elements. In most of the cases phenological models used in decision support for pest control are only based on the development rate and a one-to-one fecundity, in order to keep the population constant. Our results suggest that improvements in model performance can be obtained not only modifying the development rate functions but also considering information available on other components of the life history strategies. Data on the fecundity and mortality rate functions are available for many species, and their importance in both phenological and population dynamics models suggests to put more effort in their estimation in lab as well as in natural conditions. Among the three biological traits considered in this paper the distribution of the initial condition over the physiological age appears to be the most important one. However, this biological trait is usually not well known and only generic indications are reported in literature, with no quantitative estimation even of the interval of the distribution. 

The importance of fecundity, mortality and age distribution in the overwintering stage has to be considered for the purpose of model definition and calibration in pest management. In fact, differences of many days in the events of pest phenology can be obtained with different forms of the fecundity, the mortality or the distribution of the initial condition. This can lead to very different decision in the implementation of control strategies.

\section*{Acknowledgements}
Support by INdAM-GNFM is gratefully acknowledged by CS.\\
This research has been supported by ``Fondazione Cariplo'' and ``Regione Lombardia'' under the project: ``La salute della persona: lo sviluppo e la valorizzazione della conoscenza per la prevenzione, la diagnosi precoce e le terapie personalizzate''. Grant Emblematici Maggiori 2015-1080.\\
The authors would like to thank Tommaso del Viscio (CNR-IMATI) for technical support.

\bibliographystyle{plain}
\bibliography{bibliography}
\end{document}